\documentclass{article}
\usepackage[utf8]{inputenc}
\usepackage{anyfontsize}
\usepackage{lineno,hyperref}
\usepackage[sort&compress,semicolon,authoryear]{natbib}
\usepackage{fancyhdr}
\usepackage{enumitem} 
\usepackage{amsmath}
\usepackage{mathtools}
\usepackage{dsfont}
\usepackage{amssymb}
\usepackage{amsfonts}
\usepackage{icomma}
\usepackage{xcolor}
\usepackage{multirow}
\usepackage{graphicx}
\usepackage{pdfpages}
\usepackage{epsfig}
\usepackage{listings}
\usepackage{hhline}
\usepackage{rotating}
\usepackage{subfig}
\usepackage{afterpage}
\usepackage{float}
\usepackage{setspace}
\usepackage{hyperref}
\hypersetup{
    colorlinks,
    linkcolor={blue!50!black},
    citecolor={blue!50!black},
    urlcolor={blue!80!black}
}
\usepackage{caption}
\usepackage{longtable, tabu}
\usepackage{booktabs}
\usepackage{tabularx}
\newcolumntype{L}[1]{>{\raggedright\arraybackslash}p{#1}} 
\newcolumntype{C}[1]{>{\centering}p{#1}}
\newcolumntype{R}[1]{>{\raggedleft\arraybackslash}p{#1}} 
\usepackage{float}
\usepackage{algorithm}
\floatname{algorithm}{Procedure}
\usepackage{array}

\usepackage{calc}
\usepackage{times}
\usepackage{latexsym}
\usepackage{xr}
\usepackage{subfig}
\usepackage{makecell}

\usepackage{eurosym}  

\allowdisplaybreaks[3]
\makeatletter
\newcommand\footnoteref[1]{\protected@xdef\@thefnmark{\ref{#1}}\@footnotemark}
\makeatother

\relax
\catcode`@=11

\def\nvphantom{\v@true\h@false\nph@nt}
\def\nhphantom{\v@false\h@true\nph@nt}
\def\nphantom{\v@true\h@true\nph@nt}
\def\nph@nt{\ifmmode\def\next{\mathpalette\nmathph@nt}%
  \else\let\next\nmakeph@nt\fi\next}
\def\nmakeph@nt#1{\setbox\z@\hbox{#1}\nfinph@nt}
\def\nmathph@nt#1#2{\setbox\z@\hbox{$\m@th#1{#2}$}\nfinph@nt}
\def\nfinph@nt{\setbox\tw@\null
  \ifv@ \ht\tw@\ht\z@ \dp\tw@\dp\z@\fi
  \ifh@ \wd\tw@-\wd\z@\fi \box\tw@}

\newcommand{\iton}{ i=1,..,n }
\newcommand{\vdvafzs}{VDV~457}

\newcommand{\Var}{\operatorname{Var}}

\newcommand{\E}{\operatorname{E}}

\newcommand{\olD}{{\overline{D}}}
\newcommand{\olX}{{\overline{X}}}
\newcommand{\olM}{{\overline{M}}}
\newcommand{\hnu}{{\widehat \nu}}
\newcommand{\hM}{{\widehat M}}
\newcommand{\hX}{{\widehat X}}
\newcommand{\hD}{{\widehat D}}

\newcommand{\hp}{{\widehat p}}

\newcommand{\mys}{ \text{\texttt{s}} }
\newcommand{\myu}{ \text{\texttt{u}} }
\newcommand{\csn}{ c_{\mys  \kern-1pt \myscalebox{4.0pt}{!}{ \texttt{0} } } }
\newcommand{\csz}{ c_{\mys  \kern-1pt \myscalebox{4.0pt}{!}{ \texttt{Z} } } }
\newcommand{\cu}{ c_{\myu} }
\newcommand{\zalf}{ z_{ \myom \kern-1pt  \alpha \kern-1pt / \kern-1pt 2} }
\newcommand{\zbet}{ z_{ \myom \kern-1pt  \beta \kern-1pt / \kern-1pt 2} }
\newcommand{\myeq}{ \kern0pt \text{\texttt{=}} \kern-0pt}
\newcommand{\numin}{ \nu_{\text{min}}}

\newcommand{\nref}{ n_{\text{e}}}
\newcommand{\nrec}{ n_{\text{rec}}}
\newcommand{\mymu}{ \! \cdot \! }
\newcommand{\myom}{ 1 \kern-1pt  \text{\texttt{-}} \kern1pt   }
\newcommand{\mydp}{{d \kern-2pt \operatorname{P} \kern-1pt }} 
\newcommand{\myp}{{ \operatorname{P} \kern-1pt }}
\newcommand{\myv}{{ \kern-0pt \rvert \kern-1pt }}

\newcommand{\myscalebox}[3]{ \resizebox{#1}{#2}{$ #3 $} }

\newcommand{\rAV}{ r_{\text{A} \kern-1pt /  \kern-0.5pt \text{V}} }

\newcounter{revcomcounter}

\renewcommand{\therevcomcounter}{\arabic{revcomcounter}}

{\end{trivlist}}%

{\end{trivlist}}%

\usepackage{booktabs,tabularx}
\newcolumntype{C}{>{\centering\arraybackslash}X}

\usepackage{siunitx}

\begin{document}
\renewcommand{\figureautorefname}{Fig.}
\renewcommand{\sectionautorefname}{Section}

\title{Introducing the Partitioned Equivalence Test: Artificial Intelligence in
Automatic Passenger Counting Validation}
\author{David Ellenberger\textsuperscript{1} \and Michael Siebert\textsuperscript{1} }

\date{Received: date / Accepted: date}
\maketitle
\begin{abstract}
\begin{quote}
Automatic passenger counting (APC) in public transport has been introduced in
the 1970s and has been rapidly emerging in recent years. APC systems, like all
other measurement devices, are susceptible to error, which is treated as random
noise and is required to not exceed certain bounds. The demand for very low
errors is especially fueld by applications like revenue sharing, which is in the
billions, annually. As a result, both the requirements as well as the costs
heavily increased. In this work, we address the latter problem and present a
solution to increase the efficiency of initial or recurrent (e.g.\@{} yearly or
more frequent) APC validation. Our new approach, the \emph{partitioned
equivalence test}, is an extension to this widely used statistic hypothesis test
and guarantees the same bounded, low user risk while reducing effort. This can
be used to either cut costs or to extend validation without cost increase. It
involves a pre-classification step, which itsself can be arbitrary, so we
evaluated several use cases: entirely manual and algorithmic, artificial
intelligence assisted workflows. For former, by restructuring the evaluation of
manual counts, our new statistical test can be used as a drop-in replacement for
existing test procedures. The largest savings, however, result from latter
algorithmic use cases: Due to the user risk being as bounded as in the original
equivalence test, no additional requirements are introduced. Algorithms are
allowed to be failable and thus, our test does not require the availability of
general artificial intelligence. All in all, automatic passenger counting as
well as the equivalence test itself can both benefit from our new extension. 
\end{quote}
\end{abstract}
\newpage
\section{Introduction}\label{sec_INTRO}
\begin{figure}
  \setlength{\unitlength}{1cm}
  \includegraphics[width=0.95\textwidth]{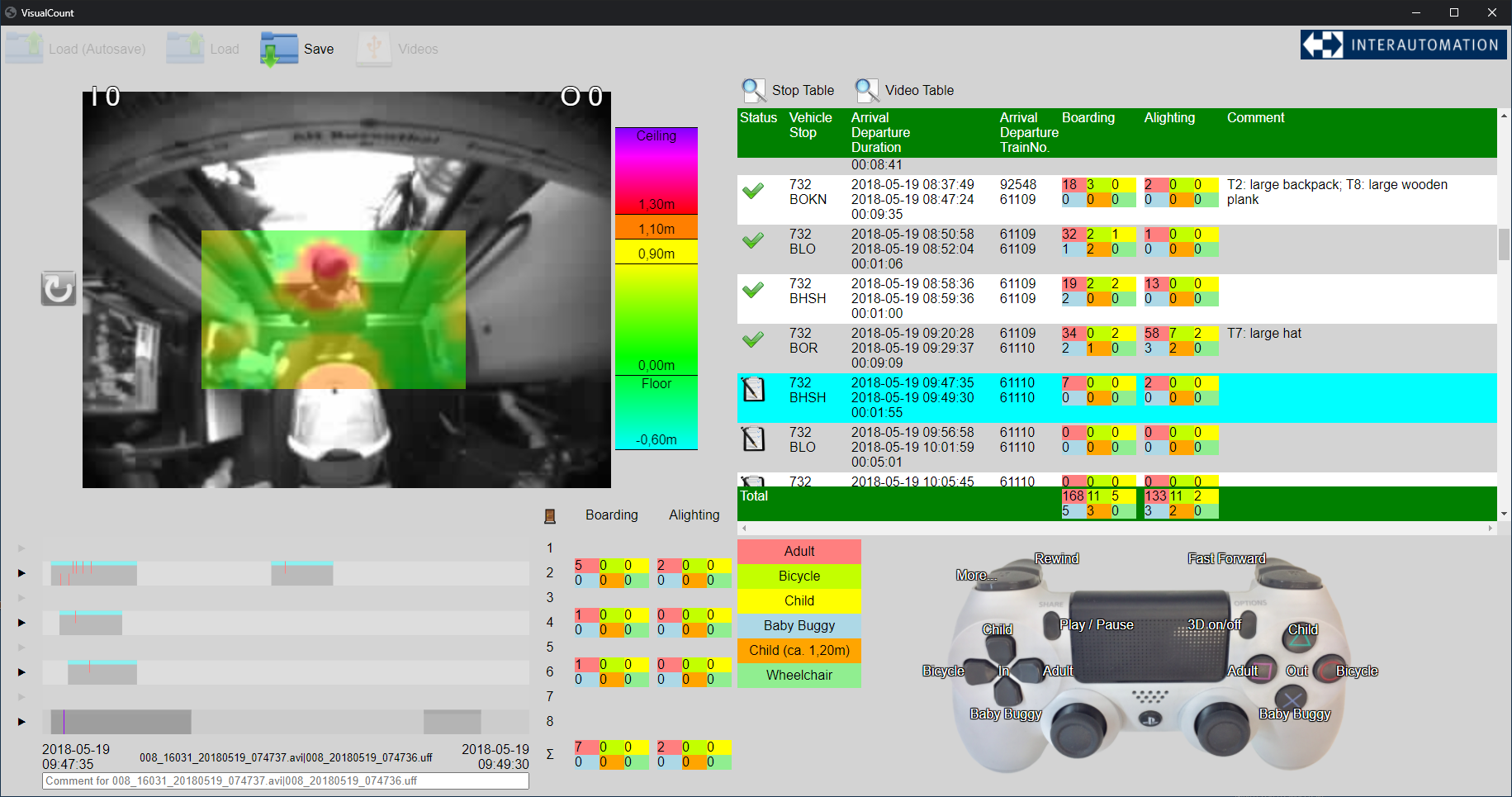}
  \centering
  \caption{Screenshot of VisualCount: a specialized video software to perform
  manual counts, which has been used to obtain our count data, compare Section
  \ref{sec_application}. The height information from 3D videos can be used to
  distinguish between children and adults, e.g.\@{} by using a critical height
  of 1.20 meters. To theses standards, the small person in the center of the
  image above would be considered as an adult, while the one to the left as a
  child. VisualCount yields a 70\% speedup over the use of a standard video
  player with a spreadsheet software (like Microsoft Excel), which reduces the
  additional costs for video evaluation from $300\%$ (the quadrupling of costs
  by accounting for the user risk in the \vdvafzs{} v2.1 vs. the old version
  v2.0) to approximately $33\%$. It runs entirely in the browser, thus can
  operate without a server by using videos from the users filesystem, which
  increases data privacy. A cloud-based operation would be possible as well.
  Performance-critial parts are implemented in C/C++ and compiled to JavaScript
  or WebAssembly with Emscripten \citep{Zakai2011}, with filesystem-call
  passthrough. This allows to determine durations, frame counts as well as
  single-frame random access to hundreds of gigabytes of video data not only
  using browser-native (like H.264 or AV1), but also general standard (e.g.\@{}
  MJPEG via FFmpeg) or specialized, custom file formats (e.g.\@{} multiple,
  losslessly compressed 3D video formats) without noticeable delay, eliminating
  almost all wait times. To ensure a seamless integration, we reverse engineered
  a proprietary video file format from one of our suppliers, who provided the
  full documentation after a technical demonstration. A gamecontroller is used
  to navigate through the video with playback or rewind speeds corresponding to
  the pressure applied to the analog buttons to obtain both slow motion for
  crowds as well as high speeds (10x and more) to skip through long idle
  timespans. Rates of 60 frames per second can be reached within the browser so
  that even in fastest fast forward mode, no boarding or alighting passenger is
  accidently skipped. A specialized video software (such as VisualCount) is
  required to reach Efficiency Level 2 and
  above.}\label{fig_visual_count_screenshot}
\end{figure}

\begin{figure}
  \setlength{\unitlength}{1cm}
  \includegraphics[height=0.25\textwidth]{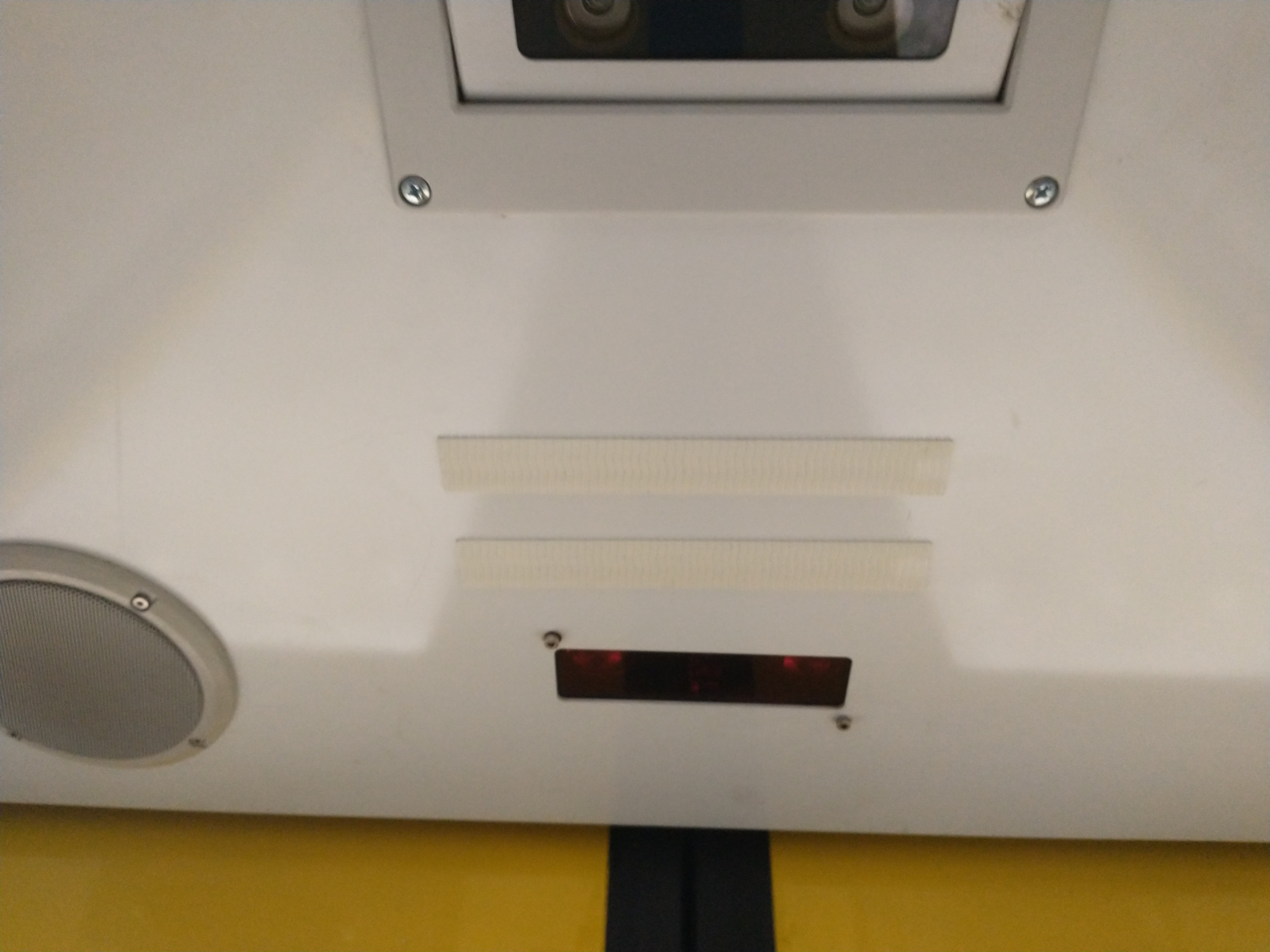}
  \includegraphics[height=0.25\textwidth]{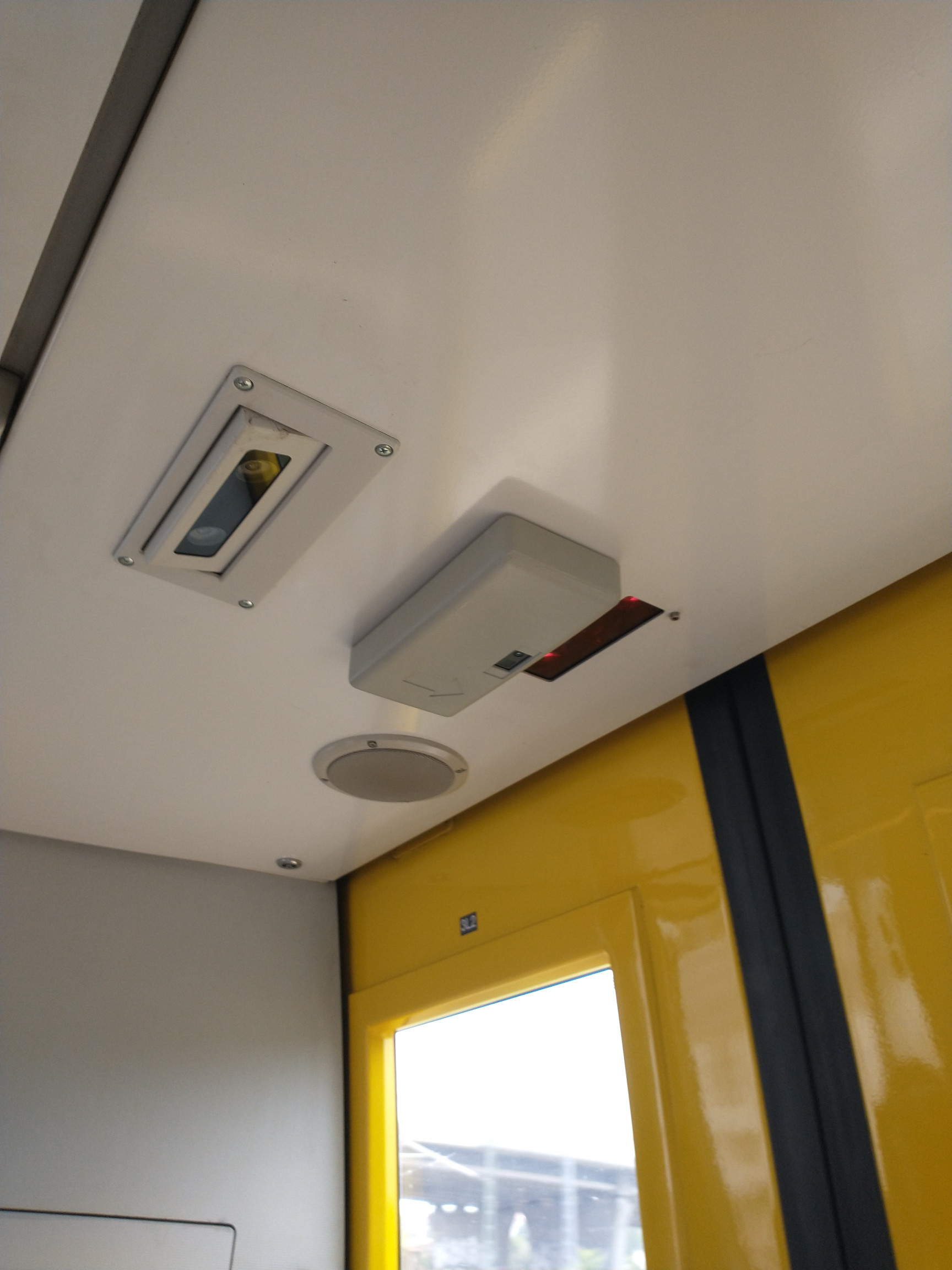}
  \includegraphics[height=0.25\textwidth]{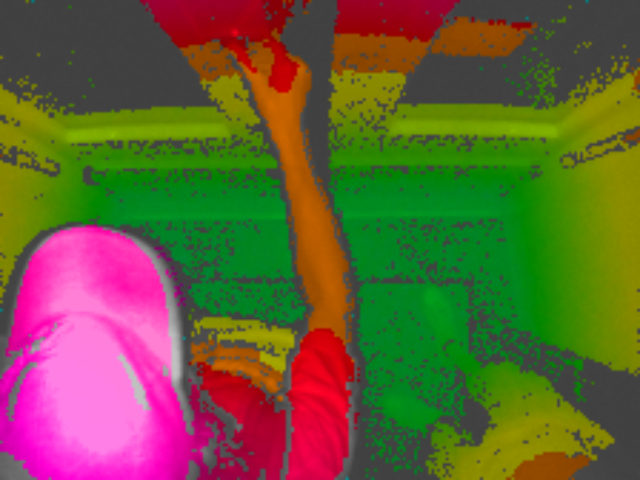}
  \includegraphics[height=0.25\textwidth]{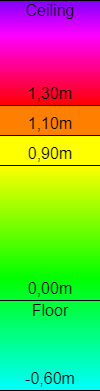}
  \centering
  \caption{Mounted with industrial velcro (first left) above doors, the Depth
  Sensing Unit DSU (second left) a portable, battery-powered 3D video recording
  device. For demo image material from VisualCount, see right image with
  colorscale.}\label{fig_dsu}
\end{figure}

Assessment of passenger counts is of paramount importance for public transport
agencies in order to plan, manage and evaluate their transit service. Over the
past three decades, automatic passenger counting (APC) systems have played an
increasingly important role in determining the number of passengers in local
public transport. They are used in the daily monitoring of operations, in
long-term demand planning, as well as in revenue sharing within transport
associations around the world. For more details and an overview of APC
development and current practice, see \citet{SiebertEllenberger2019}. Revenue
magnitudes in the billions are common in public transport \citep{Armstrong2010},
e.g.\@{} in the year 2018, total ticket revenues in Germany alone have been
12.95 billion euros \citep{vdv:PresseinformationBilanz2018}, while APC systems
are deployed worldwide. In many cases, a passenger count is the de facto
standard for public tenders or the acquisition of subsidies. The counting
quality of existing APC systems on the market has been continuously improved by
technical developments in combination with increased requirements. Nowadays, APC
systems are expected to have a maximal systematic error or bias of 1\%. This
aspect of APC validation is referred to as an accuracy of 99\%. For transport
associations in Germany, but also internationally, validation is typically
regulated by the VDV, recommendation 457 \citep{vdv457_v2_1}: In 2018, a
criterion based on the t-test was replaced by an equivalence test, which takes
the user error into account and limits it to $5\%$. This tightening of the
requirements compared to the previous test criterion has led to up to four times
larger sample sizes in the testing of the measurement accuracy, which also
quadruples the costs that arise primarily from the manual inspection of the
counting situations by comparison counting personnel. The already high cost
pressure on the comparison counting continues to increase and the need for
solutions to reduce costs increases alongside. Since the hourly wages of the
reference counters remain the same or increase in perspective, technical and
regulatory solutions are necessary.
In order to increase validation efficiency and reduce costs, we have identified
the following levels during our recent years of research (compare Table
\ref{tab_current_savings} for real world numbers):
\begin{description}
    \item[Efficiency Level 0:] Manual ride checkers that stay in the vehicle
    during its entire journey and count boarding and alighting passengers
    (and other count objects) at doors.
    \item[Efficiency Level 1:] Perform all manual counting on recorded (and
    automatically cut) (3D-)videos, which modern APC systems can acquire
    directly from the sensors, compare Table \ref{tab_current_savings}. Having a
    pool of (possibly unseen) videos available is a requirement for Efficiency
    Level 2 and 3.
    \item[Efficiency Level 2:] Increase efficiency in the evaluation of video
    material, e.g.\@{} by using application-specific software that integrates
    comparison counting and video viewing, compare Figure
    \ref{fig_visual_count_screenshot}.
    \item[Efficiency Level 3:] Reduction of the video volume to be evaluated
    manually with verifiably equal validation quality (manufacturer and user
    risk) as the current method based on the equivalence test.
\end{description}
For latter, Efficiency Level 3, the reduction of the manually evaluated video
volume, an adapted mathematical-statistical formalism is required, which is the
subject of this manuscript: We introduce and discuss the concept of this
so-called \emph{Partitioned Equivalence Test} in the second and present a
mathematical formalization third section. We show how to perform a sample size
calculation for our new method, analyse and optimize costs in the fourth
section, evaluate real world data in the fifth section and close with some
concluding remarks and future prospects in the last section. In the following we
assume that a single \emph{(3D-)video} corresponds to a \emph{door opening phase
(DOP)}, i.e.\@{} it shows a view from inside the vehicle that allows to see an
entire indoor door area (compare Figure \ref{fig_visual_count_screenshot}) and
Figure \ref{fig_dsu}) from door opening to door closing:
\begin{enumerate}
  \item Door opening: initially, the door is (almost) completetly closed, so
  that no passenger (or countable object) might pass.
  \item All the events that happen while the door is open, e.g.\@{} boarding and
  alighting passengers.
  \item Door closing: at the end of the video, the door is (almost) completely
  closed again so that no passenger (or countable object) might pass.
\end{enumerate}
A commonly used unit for (automatic) passenger counting so far has been the
\emph{Stop Door Event} (SDE), which corresponds to a video of an entire door
during an entire stop: e.g.\@{} if a door (e.g.\@{} door 2) opens and closes
again 3 times during a stop, 3 DOP and one SDE is generated. However, nowadays
APC systems allow to record DOP, e.g.\@{} by accessing the corresponding door
opening signal, which reduces the amount of video data to be manually evaluated
considerably and upfront, compare Table \ref{tab_current_savings}. From an
analytical point of view, the standard deviation can be estimated more reliably,
since door opening phases are the smaller statistical unit than stop door
events. In the following we therefore use door opening phases. The sample size
calculation may still be carried out on stop door events e.g.\@{} for comparison
purposes, see Section \ref{sec_sample_size_calculation}.
\begin{table}
  \setlength{\unitlength}{1cm}
  \begin{tabularx}{\linewidth}{*{1}{l} *{7}{C}}
\toprule
& Efficiency Level 0 & \multicolumn{3}{c}{Efficiency Level 1} &
Efficiency Level 2 & \multicolumn{2}{c}{Efficiency Level 3} \\
& Ride Checkers in Vehicle & Vehicle Stop & Stop Door Event & Door Opening Phase
& Specialized Video Software & Manual Partitioned Equiv. Test & Algorithmic
Partitioned Equiv. Test \\
\midrule
Duration & 38:43:26 & 15:52:20 & 05:45:11 & 02:48:15 & 00:50:29 & 00:40:23 &
00:25:14\\
\\
\multicolumn{4}{l}{Manual Partitioned Equiv. Test} & & & & 37,50\% \\
\multicolumn{4}{l}{Specialized Video Software} & & & 20,00\% & 50,00\% \\
\multicolumn{2}{l}{Door Opening Phases} & & & & 70,00\% & 76,00\% & 85,00\% \\
\multicolumn{2}{l}{Stop Door Events}
& & & 51,26\% & 85,38\% & 88,30\% & 92,69\% \\
\multicolumn{2}{l}{Vehicle Stop}
& & 63,75\% & 82,33\% & 94,70\% & 95,76\% & 97,35\% \\
\multicolumn{2}{l}{Ride Checkers}
& 59,01\% & 85,14\% & 92,76\% & 97,83\% & 98,26\% & 98,91\% \\
\bottomrule
\end{tabularx}
\caption{Savings w.r.t.\@{} different video material granularity for an urban
railway transport system. The original video data (38:43h) has been recorded
with a GoPro connected to an external battery and represents the original ride
checker approach. From the actual timetable, real arrival and departure times
were known, so the original video could automatically be cut into smaller pieces
accordingly. From an APC system installed in the vehicle, the door opening
phases were known as well and utilized in a similar way. Stop door events is a
granularity in between vehicle stops and door opening phases: all videos with no
door opening have been removed from the vehicle stop set, which is around 50\%,
since in most cases only one side of the vehicle doors open at stops.
Specialized video software (e.g.\@{} VisualCount, compare Figure
\ref{fig_visual_count_screenshot}) savings are 70\% according to our studies for
the reason that users fast-forward through idle sequences. Manual and
algorithmic partitioned equivalence tests have been accounted for with 20\%
resp.\@{} 50\% savings. All in all, Efficiency Level 3 reduces the manual effort
by around 99\% over Efficiency Level 0.}
\label{tab_current_savings}
\end{table}
\section{Partitioned Equivalence Test Concept} \label{sec_CONCEPT}
As mentioned in Section \ref{sec_INTRO}, in order to achieve a reduction in
certification costs beyond Efficiency Level 2 (the use of more integrated hard-
and software solutions) a further reduction in video volume to be evaluated
manually is necessary.

But how can this reduction be achieved methodically? Here, we initially had the
idea of using an additional algorithm to pre-classify the videos according to
the certainty or \emph{safeness} of a correct count by the APC system, so that a
manual count is only necessary on material with a considered highly uncertain or
\emph{unsafe}. The challenge would have been to formulate rules for the general
approval of such a classification system and to develop an algorithm that is
powerful enough to meet the formulated requirements. However, as it turned out
in the course of the project, this approach has unmanageable implications: it
either requires the construction of a kind of superior, infallible algorithm,
which can, with unlimited certainty, identify the incorrectly counted videos.
Alternatively, if that algorithm had been fallible, the \emph{ground truth}
would have had to be redefined: currently, a count value is considered correct
if it was generated by (at least) two manual counters and verified a third
manual counter -- the supervisor -- in case the first two counters differ
\citep{vdv457_v2_1}. An attempt to redefine this current ground truth would have
raised both ethical as well as technical questions
\citep{lake_ullman_tenenbaum_gershman_2017} and would definitely have lead to
unfruitful, never ending discussions in the foreseeable future. Facing these
challenges, we finally changed our approach to include a sample in the material
that the algorithm has classified as \emph{not necessary to be manually viewed}
or \emph{safe}. Here the question arises whether savings can be achieved in this
way at all, but such a procedure is at least technically feasible by today's
standards and also methodically sound. We have therefore continued our
investigations.

For the implementation of Efficiency Level 3 (reduction of the video volume to
be evaluated manually with verifiably equal validation quality, compare Section
\ref{sec_INTRO}), a mathematical-statistical formalism is required which
satisfies the following requirements:
\begin{enumerate}
\item Automatic passenger counting and the VDV 457 influence the worldwide
distribution of revenue in public transport, which means that all changes,
especially in the statistical inference, are critical. This results in the
following fundamental requirements:
    \begin{enumerate}
    \item The cost savings should be relevant enough to make the adaptation of
    an existing validation procedure legitimate.
    \item For a well-founded decision, an analytical derivation of the new
    statistical method must be possible.
    \item The cost savings should be achievable without changing the definition
    of the ground truth.
    \end{enumerate}
\item Compared to the equivalence test, the new test should not place any
    additional requirements on the statistical distribution of counting errors.
\item If the parameters of the new test are selected so that the entire sample
    is counted, the new test should correspond to the previous test, i.e.\@{}
    the equivalence test.
\item As in the equivalence test, it should always be ensured that the user risk
    is not greater than a specified limit (currently max.\@{} $5\%$) and a
    specialized sample size estimation should lead to a controllable adaptation
    of the manufacturer risk.
\item It would be beneficial if no special software is required to determine the
    result of the test, i.e.\@{} an evaluation itsself should be possible with
    commonly used spreadsheet software.
\end{enumerate}
We have created a test procedure that statisfied all the above mentioned
requirements, the \emph{Partitioned Equivalence Test}. The idea is as follows:
First, divide (or \emph{partition}) the video material to be evaluated into two
parts: one part comprises the unsafe door opening phases, i.e.\@{} all videos
where a miscount of the APC is suspected. This entire so-called \emph{unsafe
partition} is counted manually according to the current procedure, i.e.\@{} by
(at least) two persons and a supervisor count in case the first two differ. In
the other part, the so-called \emph{safe partition}, there is only a
\emph{relatively small} manual sample count. The \emph{partitioned} equivalence
test is now able to merge the comparison counts of the two partitions again and
to create a common confidence interval, on which a regular equivalence test can
be carried out. The partitioned equivalence test is performed in four steps,
compare Figure \ref{fig_pe_concept}:
\begin{figure}[H]
  \setlength{\unitlength}{1cm}
  \includegraphics[page=1,width=0.48\textwidth]{%
  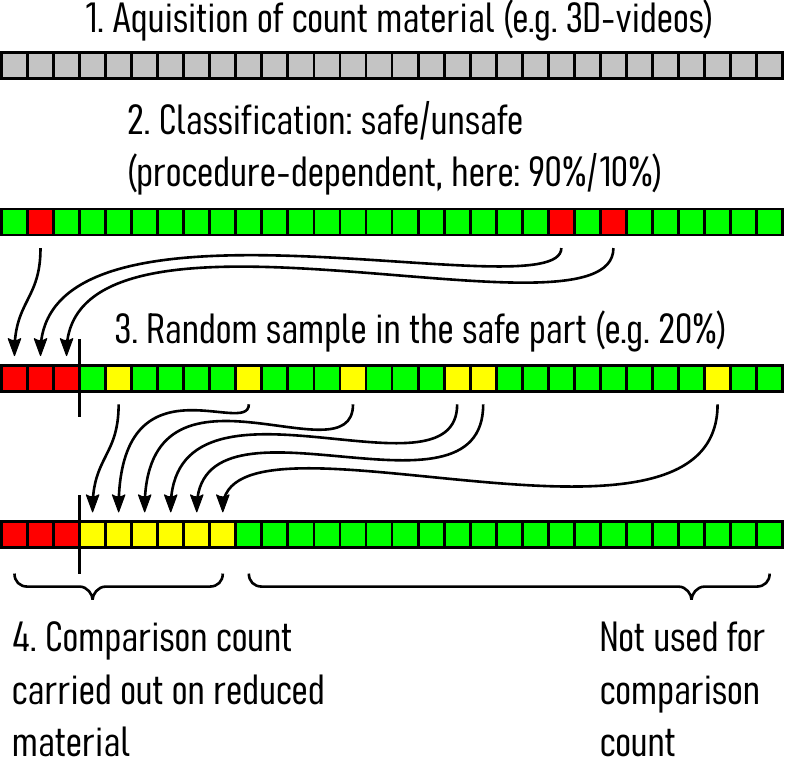}
  \centering
  \caption{Schematic representation of the partitioned equivalence
  test.}\label{fig_pe_concept}
\end{figure}
\begin{enumerate}
    \item Acquisition of the counting material (e.g.\@{} 3D videos)
    \item Creation of a \emph{partition} by classification into safe and unsafe
    door opening phases
    \item Selection of a random sample in the safe partition
    \item Carrying out the comparison count on the reduced material
\end{enumerate}
\section{Statistical Model} \label{sec_MODEL}
As a starting point we take the model as in \citep{SiebertEllenberger2019}: Let
$\Omega_0=\{\omega_i\}, i=1,\ldots, \infty$ be the statistical population of
\emph{door opening phases} (\emph{DOP}), which are used to summarize all
boarding and alighting passengers at a single vehicle (bus, tram, train) door
during a door opening of that door. Further, let $\Omega=\{\omega_{i_j}\},
i_j\in\{1,\ldots, \infty \}, \ j\in\{1,\ldots, n\}$ be a sample, which consists
of $n$ either randomly or structurally selected door opening phases (e.g.\@{} by
a given sampling plan). We use the notation $n=n_e$ when the sample size was
planned according to the equivalence test. Let $M_i$, $i\in\{i,\ldots,n\}$ be
the \emph{manual count}, $K_i$, $i\in\{i,\ldots, n\}$ be the \emph{automatic
count} of boarding passengers made by the APC system and $X_i = K_i-M_i$,
$i\in\{i,\ldots, n\}$ the differences among them. For the partitioned version of
the equivalence test yet to be introduced, it does not suffice to obtain the
manual count by ride checkers, (3D-)videos of door opening phases are mandatory:
We require a supply of (possibly unseen) footage, which can subsequently be used
for counting through multiple manual sightings of single videos, favourably 3D
depth data or 2D videos with additional, lower resolution 3D
information. 3D depth data allows to estimate the persons heights, which
is used to distinguish between adults and children and which is especially
relevant for revenue sharing, since tickets for children are typically sold at a
reduced price. To create the ground truth counts, only manual counts may be
used: at least two with an additional supervisor count in case of differences.
Alighting passengers (or other objects) are counted as well and results apply
analogously, but without loss of generality we only consider the boarding
passengers. Let $\overline{M}=\frac{1}{n}\sum_{i=1}^{n} M_i$ be the average
manual boarding passenger count. We consider the random variables
\begin{align}
    D_i := \frac{K_i-M_i}{\overline{M}}\ \ ,
\end{align}
which we call \emph{relative differences} being the differences of the
automatically and manually counted boarding passengers relative to the average
of the manually counted boarding passengers. The average $\olD:=\frac{1}{n}
\sum_{i=1}^{n} D_i$ is the statistic of interest which is used in the
equivalence test. The expected value $\mu_{\Omega}:=E(\olD)=\frac1{n} \kern-1pt
\sum_{i=1}^n \kern-2pt E(D_i)$ is the \emph{actual systematic error} of an APC
system, since it can systematically discriminate participants of the revenue
sharing system. It could also be referred to as \emph{bias} of the measurement
device \citep[see e.g.\@{}][]{Nielsen2014} or as \emph{statistical distortion}
\citep[see e.g.\@{}][]{vdv457_v2_1}. Let $ \nu_i^2 := E(D_i \kern-2pt -
\kern-2pt \mu_{\Omega} )^2$ be the quadratic error of the $D_i$ relative to the
expected value $\mu_{\Omega}$. It corresponds to the \emph{variance} of $D_i$ in
the case of $E(D_i) \! = \! \mu_{\Omega}$ and $\nu^2 := \frac{1}{n} \sum_{i=1}^n
\nu_i^2$, which is the mean square error of $D_i$. The square root $\nu =
\sqrt{\nu^2}$ corresponds to the definition of the \emph{standard deviation},
while $ \hnu^2  := \! \frac1{n-1} \kern-1pt \sum_{i=1}^n (D_i \kern-1pt -
\kern-1pt \olD)^2  $ is the empirical variance estimator for $\nu^2$.

To test for \emph{equivalence}, one wants to show that observed differences are
within certain bounds, as opposed to complete equality. We here use the
hypotheses and error types as commonly defined for equivalence testing, also
sometimes referred to as the \emph{two one-sided tests} (TOSTs) procedure
\citep{schuirmann1987}. As an alternative approach, the equivalence
test can be derived directly from the two-tailed t-test under certain simple
assumptions, i.e., that the parameters are induced (or simply exchanged) from
those of the t-test \citep{SiebertEllenberger2019}.

Thus, the hypotheses are \citep{julious2004}
\begin{align}
H_0&:\ \text{There is a (relevant) systematic APC measurement error}(|\mu_{\Omega}| \ge \Delta )\\
H_1&:\ \text{There is no (relevant) systematic APC measurement error }(|\mu_{\Omega}| < \Delta)\ \ .
\end{align}
We define $\Delta$ to be the equivalence margin and the relevant errors for the
equivalence test with $\alpha$ referring to (half) the risk of the user and
$\beta$ to the risk of the device manufacturer. We will consider two-sided $1 -
2\alpha$ confidence intervals where $\alpha$ is commonly chosen to be $2.5\%$.
The test criterion to be evaluated is 
\begin{gather}
|\olD| \le \Delta - z_{1-\alpha}\ \frac{\hnu}{\sqrt{n}}\ \ .
\label{eq_aequivalenz_orig_inequality}
\end{gather}
Sample size estimation for an equivalence test defined this way is given by
\citet[][]{julious2004} as:
\begin{gather}\label{eq_sample_size_eq}
n = \bigl(z_{1-\beta/2}+z_{1-\alpha}\bigr)^2\ \frac{\nu^2}{\Delta^2}\ \ .
\end{gather}
To develop the partitioned equivalence test as outlined, one uses an already
existing classification into safe and unsafe door opening phases (DOP) on the
total sample $n_{\text{rec}}$ of recored videos ($n_e \leq n_{\text{rec}}$),
which results in two partitions. This classifcation is given by $W_i \, , \;
\iton$ into \emph{safe} ($W_i \! = \! \mys$) and \emph{unsafe} ($W_i \! = \!
\myu$) door opening phases (DOP). To improve readability, we use the character
placeholders $\mys$ and $\myu$ but any indicator will suffice. Further, let
$p_\mys \, ,  \; p_\myu \kern-2pt = \kern-2pt \myom  p_\mys$ be the likelihood
$p_\mys$ of a DOP to be classified as \emph{safe} and the (counter) probability
$p_\myu$ of a DOP to be classified as \emph{unsafe}. Then $W_i$ is accordingly
Bernoulli distributed: $ W_i \sim \myp^W \kern-5pt = \! \operatorname{Bin}(1,
\kern1pt p_\mys)$. Let $N_\mys \, , \; N_\myu \! = \! N \kern-3pt - \kern-3pt
N_\mys$ be the total number of safe and unsafe DOP and $\hp_{\mys} \kern-1pt =
\kern-1pt N_\mys \kern-1pt / n$ be the frequency of safe DOP. Analogously, let
$\mu_{\mys} \, , \; \nu_{\mys} \, , \; \mu_{\myu} \, , \; \nu_{\myu}$ be the
corresponding parameters for $\mu$ and $\nu$ on the respective partitions of the
safe DOP and the unsafe DOP, explicitly $\mu_{\mys} = {\textstyle
\frac1{N_{\mys}} \kern-3pt \sum_{k=1}^{N_{\mys} } } \mu_{\mys k} $, $\mu_{\myu}
= {\textstyle \frac1{N_{\myu}} \kern-3pt \sum_{k=1}^{ N_{\myu} }} \mu_{\myu k} $
and $\nu^2_{\mys} = {\textstyle \frac1{N_{\mys}} \kern-3pt \sum_{k=1}^{N_{\mys}}}
 \nu^2_{\mys k} $ and $\nu^2_{\myu} = {\textstyle \frac1{N_{\myu}} \kern-3pt
\sum_{k=1}^{N_{\myu} } } \nu^2_{\myu k}$. Let $q$ be the so-called \emph{counted
quota} of the safe partition and $1-q$ the fraction not to be counted, i.e.\@{}
skipped.

Different procedures, the so-called \emph{use cases}, to obtain such partitions
are described when discussing application cases in Section
\ref{sec_application}. In this context, useful additional information, such as
video data and stop characteristics, can be used in a meaningful way to achieve
a classification $W_i$, $\iton$. This implies that, in the following,
conditional distributions (and parameters) on \emph{safe} DOP and on
\emph{unsafe} DOP are to be considered, e.g. for $M_i \, \myv W_i $. A critical
part of the use cases is the cost control, which is introduced in Section
\ref{sec_COST}: taking into account basic costs $\csn$, as well as manual
counting costs $\csz$ for safe DOP and the combined costs for unsafe DOP $\cu$.

\begin{figure}[H]
   \setlength{\unitlength}{1cm}
    \includegraphics[page=1,width=0.78\textwidth]{%
      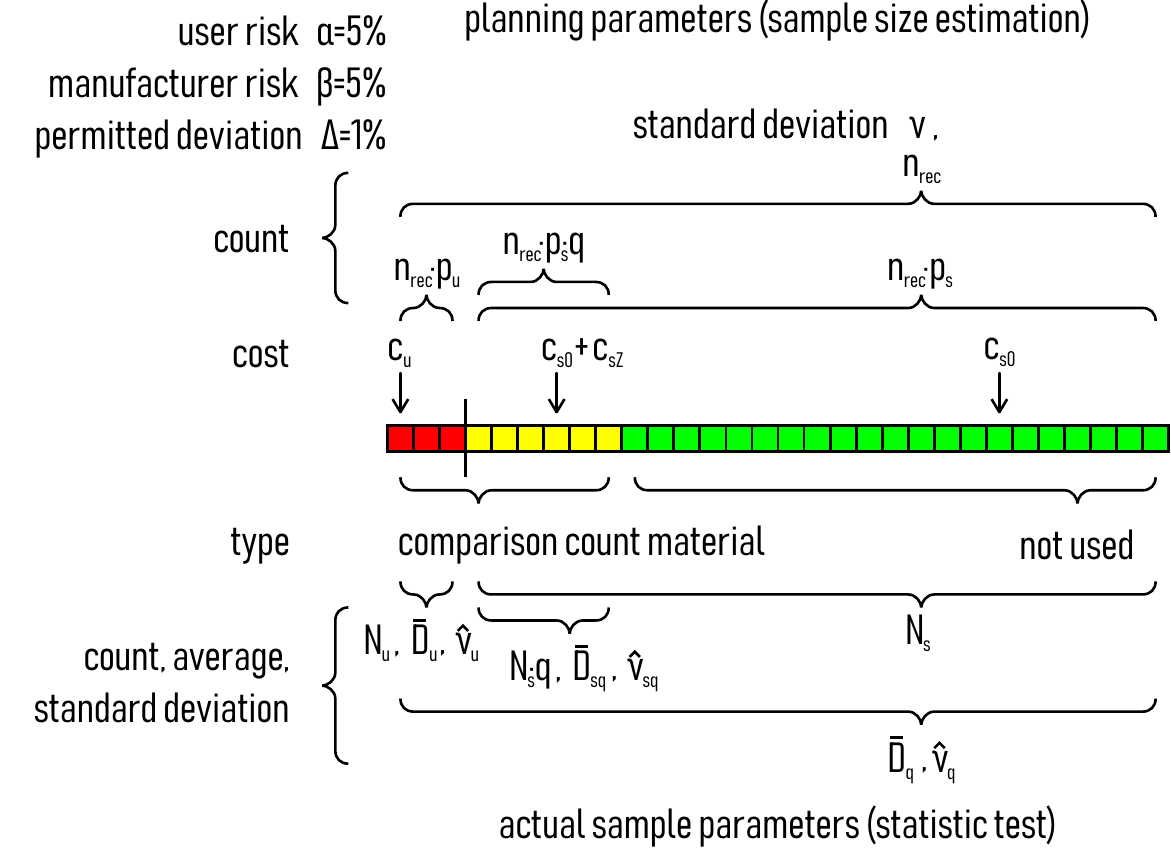}
    \centering
    \caption{Schematic representation of the parameters of the partitioned
    equivalence test.}\label{fig_pe_concept_params}
\end{figure}

For a schematic representation, compare Figure \ref{fig_pe_concept_params} of the
current as well as the newly introduced parameters. The alighting passengers are
not explicitly mentioned, as they are handled analogously to the boarding
passengers.
Our general statistical model is thus based on two strata that are defined by
the two partitions. Since the classification into a safe and an unsafe partition
can be entirely arbitrary, our model used stems from a general mixture model. A
two dimensional mixture model can described as a hierarchical model consisting
of the random classification as mixture component, see e.g.\@{}
\citet{Mclachlan1988} or for Gaussian mixture models \citep{Reynolds1993}.
Methods described in the statistical literature usually consider estimating the
parameters of the mixture components. In our case, the classification is known
and estimation of this component is thus not required. The estimation here is
only the weighted recombination of both partitions. The case of combining data
from different partitions (i.e.\@{} sources) resembles a fixed-effects meta
analysis, but is different in two aspects. First, weights are not determined
solely by sample size or standard error but rather by an artificial weighting
scheme that is induced to avoid any overrepresentation that might be introduced
by any (free choice) $q$. Secondly, the weights are not fixed but are dependent
on the randomness of the classification, i.e. to be considered a random
variable. In the following two sections we will introduce methods that address
both aspects.
\subsection{Estimation of the expected manual count}
As a first step for all further calculations with the partitioned equivalence
test, it is necessary to (indirectly) estimate the mean number of boarding
passengers in the total sample $\olM$ because in the safe partition the ground
truth is only determined, i.e.\@{} manually counted, for a proportion $q$. Thus,
in addition to the information of the measurement error $X_i$, the
non-comparison counted data also lacks the information of the actual $M_i$. The
average boarding number is needed for the definition of $D_i$ and has to be
estimated, since the mean value $\olM$ is not available. We here use the
leave-q-out estimator (see Appendix \ref{sec_KI} for details). With 
\begin{gather}
  \overline{M}_\myu  := {\textstyle \frac1{N_{\myu}} \kern-3pt
  \sum_{k=1}^{ N_{\myu} } }  M_{\myu k}
  \qquad \qquad  \hM_{\mys q} := {\textstyle \frac1{N_{\mys}} \kern-3pt
  \sum_{k=1}^{ N_{\mys} } } M_{\mys k}
  \cdot  Z_k  \kern-1pt / \kern-1pt q
\end{gather}
we obtain the estimator
\begin{align}
 \hM_q &:= \hp_\mys \mymu \hM_{\mys q} + \hp_\myu \mymu \olM_\myu  \\
     &\kern3pt = \textstyle \frac1n \! \! \left( \sum_{k=1}^{N_\myu} M_{\myu k} +
      \sum_{k=1}^{N_\mys} M_{\mys k}
     \! \cdot  \!  Z_k  \kern-1pt / \kern-1pt q \right)
\end{align}
for the average number of boarding passengers. This value can now be used
instead of $\olM$ when calculating $D_i$:
\begin{gather}
  D_i := \frac{X_i}{\hM_q}
\end{gather}
and in the case of a full manual count $q=100\%$, the definition of $D_i$ is
identical to the definition of the regular equivalence test because
$\hM_{q=100\%} \! = \! \olM $.
\subsection{Estimation of APC bias}
Analogously to the estimation of the expected manual counts and with the the
expected values of safe $\mu_\mys$ and unsafe $\mu_\myu$ DOP, a similar result
regarding the expected value holds (see Appendix \ref{sec_KI} for details with
the random variables $X_{\myu k}$ being the relative differences $D_{\myu k}$):
\begin{gather}
    \E( \olD ) =  p_\mys \mymu \mu_\mys  + p_\myu \mymu  \mu_\myu
\end{gather}
Methods are now needed to obtain a range-preserving confidence interval if only
a portion specified in advance (for example $20 \%$) of the safe DOP is to be
counted. A random selection of $q \mymu N_\mys$ safe DOP is formed which are to
be included in the final evaluation -- i.e.\@{} for which the actual ground
truth must be determined. The mean value serves as the estimator for $\mu_\myu$.
\begin{align}
    & \overline{D}_\myu  := {\textstyle \frac1{N_{\myu}} \kern-3pt
       \sum_{k=1}^{ N_{\myu} } } \,  D_{\myu k}
    \intertext{ and for $\mu_\mys$ a similar estimator can be obtained,
    which, however, only uses the quota $q$. The $Z_j$ indicate herby,
    if the DOP was randomly selected ($\myeq 1)$ or not ($\myeq 0)$:}
    & \hD_{\mys q} :=  {\textstyle \frac1{N_{\mys}} \kern-3pt
    \sum_{k=1}^{ N_{\mys} } } \,  D_{\mys k}
    \cdot  Z_k  \kern-1pt / \kern-1pt q
    \intertext{From the different estimators one can now generate some kind
    of composite estimator for $\mu$:}
    & \hD_q := \hp_\mys \mymu \hD_{\mys q} + \hp_\myu \mymu \olD_\myu
    \label{eq_leaeveQoutEST}\\
    & \kern17pt = \textstyle \frac1n \! \left(  \sum_{k=1}^{N_\myu} D_{\myu k} +
     \sum_{k=1}^{N_\mys} D_{\mys k} \! \cdot  \!  Z_k  \kern-1pt / \kern-1pt q \right)
\end{align}
All values $D_{\myu j}$ where $Z_j$ is zero are therefore no longer needed for
calculations. Then
\begin{align}
    \E (\hD_q) &= \textstyle \frac1n \cdot \E \! \left( \kern-1pt \frac{N_\mys}{N_\mys}
    \sum_{k=1}^{N_\mys} D_{\mys k} \cdot  Z_k  \kern-1pt / \kern-1pt q
    \, + \, \frac{N_\myu}{N_\myu} \sum_{k=1}^{N_\myu} D_{\myu k} \! \right) \\
    &= \textstyle \frac1n \cdot \left( p_{\mys} \mymu n \cdot \mu_{\mys} \mymu \E(Z_1)
      \kern-1pt / \kern-1pt q   + p_{\myu} \mymu n  \cdot \mu_{\myu} \right) \\
    &= p_{\mys} \mymu \mu_{\mys} + p_{\myu} \mymu \mu_{\myu} = \mu
\end{align}
and thus $\hD_q$ is an unbiased estimator which can be calculated on a reduced
dataset. Using the results of Appendix \ref{sec_KI},
\begin{align}
    \Var (\hD_q) &=  \textstyle \frac1{n}   \bigl( \,  p_{\mys} \kern-1pt \mymu \kern-1pt
    \nu^2_{\mys}
    \kern-1pt / \kern-1pt q   \,
    + (\myom p_{\mys}) \mymu \nu^2_{\myu} \;
    + \;  (\mu_{\mys} \kern1pt \text{\texttt{-}} \kern1pt \mu_{\myu}  \kern-1pt )^2
    \kern-1pt \mymu p_{\mys} \kern-2pt \mymu \kern-1pt (  \kern-1pt \myom p_{\mys}) \bigr)
    \label{eq_varest}
\end{align}
holds for the variance of the estimator. It should be noted that the variability
of the uncertainty classification (as a random variable) $(\mu_{\mys} \kern1pt
\text{\texttt{-}} \kern1pt \mu_{\myu} \kern-1pt)^2 \kern-1pt \mymu p_{\mys}
\kern-1pt \mymu \kern-1pt (  \kern-1pt \myom p_{\mys} \kern-1pt ) $ may account
for a substantial part of the variance if the expected values  $\mu_{\mys}$ and
$\mu_{\myu}$ should be very different. If the uncertainty is assumed to be
fixed, this part would not be taken into account, which would lead to too narrow
confidence intervals. This yields the (asymptotic) confidence interval:
\begin{align}
    \left[ \hD_q \pm z_{1- \alpha /2} \mymu \Var(\hD_q) \right]
\end{align}
where the unknown parameters $p_{\mys}$, $\mu_{\mys}$, $\mu_{\myu}$,
$\nu^2_{\mys}$, $\nu^2_{\myu}$ can be replaced by the empirical variance
estimators as plugin estimators. Details on variance estimation are described in
the following section.
\section{Sample size calculation}\label{sec_sample_size_calculation}
For the sample size calculation, we introduce a type II error adjustment, a
minimal (relative) standard deviation and consider costs.

\subsection{Type II error adjustments}\label{sec_type2_adjustments}
The required sample size calculation for the equivalence test in the case
without partitioning can be obtained from \citet{SiebertEllenberger2019}. With
the partitioned equivalence test, however, the challenge arises that a reduction
of the sample size (initially) increases the manufacturer risk. In order to
compensate for this increased risk, an adjustment of the sample size is
necessary, which we call the \emph{recorded size}. The sample size calculation
of the partitioned equivalence test can be directly derived from that of the
previous conventional equivalence test, with the previous standard error $\nu \!
/ \!\sqrt{n}$ replaced by the one determined in equation \eqref{eq_varest}. This
results in the following formula:
\begin{align}
    n &= ( \zalf \kern-1.5pt + \kern-1pt  \zbet  \kern-1pt )^2  \cdot \frac{
        p_{\mys} \kern-1pt \mymu \nu_{\mys}^2  / \kern-1pt q \,
    + \; p_{\myu} \kern-1pt \mymu \nu_{\myu}^2  \;
    + \; p_{\mys} \kern-1.5pt \mymu \kern-0.5pt p_{\myu}  \kern-1.5pt \mymu \kern-1pt
     (\mu_{\mys} \kern1pt \text{\texttt{-}} \kern1pt \mu_{\myu}  \kern-1pt )^2}{ \Delta^2 }   \ \ .
\end{align}
The following applies to the variance on the total sample in case of
partitioning (see Equation \ref{eq_varest} summed over the full sample in the
case $q=1$)
\begin{align}\label{eq_nu_def}\nu^2 &= p_\mys
    \kern-1.5pt \mymu \nu_{\mys}^2  \kern2pt \texttt{+} \kern2pt p_\myu \kern-1.5pt
    \mymu \nu_{\myu}^2  \kern2pt \texttt{+} \kern2pt p_\mys \kern-1.5pt \mymu p_\myu
    \kern-1.5pt \mymu \kern-0.5pt ( \mu_{\mys}  \kern-1pt \texttt{-}  \mu_{\myu}
    \kern-1pt )^2\ \ ,
\end{align}
which allows the following
\begin{align} \label{eq_ssc}
    n & = ( \zalf \kern-1.5pt + \kern-1pt  \zbet  \kern-1pt )^2  \cdot \frac{1}{
       \Delta^2 }\biggl[ p_{\mys} \nu_{\mys}^2\Bigl(\frac{1}{q}-1\Bigr)+\nu^2\biggr]    \ \
\end{align}
simplified representation. Compared with the calculations for the previous
equivalence test, this results in
\begin{align}
    n =\nrec = \nref \kern-1pt \cdot \biggl[ \frac{ p_{\mys} \nu_{\mys}^2}{\nu^2}
    \Bigl(\frac{1}{q}-1\Bigr)+1\biggr] \qquad \text{with}
    \qquad \nref = (\zalf+\zbet)^2 \frac{\nu^2}{\Delta^2}\ \ ,
\end{align}
i.e.\@{} since $0<q\le 1$ and thus $(1/q-1) \ge 0$ and $p_{\mys}\ge 0$ it can
always be viewed as multiplication of $\nref$ by a factor $\ge 1$ and thus is an
(apparent) enlargement of the sample. In fact, the partitioned equivalence test
initially only increases the \emph{recorded sample}, i.e.\@{} more comparative
video footage, of which only a part is manually counted. The added value of
first recording more and then omitting material again during the evaluation is
that a higher proportion of unsafe events can be sighted compared to the
original equivalence test. This property enables the partitioned equivalence
test, after optimizing the costs, to make a more precise statement about the
systematic error (or bias) of the counting error with less overall effort than
the original equivalence test.
\subsection{Minimal Standard Deviation}\label{sec_NUMIN}
At very small sample sizes, it is difficult to reliably estimate any parameters,
including the standard deviation, which is a problem known for example as
\emph{small-sample bias} \citep{hummel2005}. This affects the equivalence test
in general, but special attention must be paid to the partitioned equivalence
test in particular, since sample sizes in the safe partition can be very small
and errors very rare.
To better understand the implications for practice, we ran simulations with real
world APC system count errors (for more details, see Appendix
\ref{sec_VarianceSmall}) and in some scenarios the user risk $\le \alpha$ indeed
cannot be ensured for inappropriately small sample sizes.
However, this is more of a hypothetical problem, since for very small sample
sizes the overall chance to pass the test is below 15\%, independent of the
actual error $\olD$ of the APC system (even for $\olD=0$) and this can therefore
not be made a sustainable business model for any APC manufacturer. From an
authority's perspective, this is still undesirable because the chance of
approving an APC system due to a poorly designed validation is greater than the
user risk $\alpha$ implies.
For any (partitioned) equivalence test, a \emph{minimal (relative) standard
deviation} $\numin$ can be introduced as a restriction, which solves this
problem: if $\hat{\nu}<\numin$ is encountered anywhere it is replaced by
$\numin$ (in our case $\hat{\nu}_{\mys q}<\numin$ or $\hat{\nu}_{\myu}<\numin$).
Surprisingly, this allows the partitioned equivalence test to operate even more
theory-compliant than the original equivalence test. For the latter, introducing
a low $\numin$ like $\numin=3\%$ has almost no effect, since there is no
separtation into safe and unsafe videos and thus $\hnu\le\numin$ only very
rarely occurs.
\subsection{Cost Management}\label{sec_COST}
To minimize costs (see also Appendix \ref{sec_CostsExamples}), we introduce
$\csn$, the \emph{basic costs} of a safe DOP. These include, for example,
marginal costs of video recording, marginal costs of data preparation and
execution of the additional algorithms, as well as costs due to time delays in
carrying out further comparison procedures. Further, let $\csz$ be the costs
incurred for a manual comparative counting of safe DOP so that it can be
considered as a ground truth. The video data must be viewed by at least two
human counters and by a supervisor in case of conflicts. These so-called
counting costs are mainly composed of the personnel costs of the counters and
the costs due to time delays in the testing process. Finally, $c_{\myu}$ is the
(average) combined cost of the unsafe DOP, i.e. consisting out of to basic costs
and the counting costs, since both are always carried out here.
With the definition of the above partial costs, the total project cost of a
validation process can be approximated by the following function:
\begin{gather}
  \operatorname{Cost}(n,q) =  \nrec \cdot \big( p_\myu \! \! \cdot \!
  c_\myu + p_\mys \!  \cdot  \! ( \csn \! + q \! \cdot \! \csz) \! \big) \ \ .
\end{gather}
This approximation takes into account taking into multiplicity in which the
partial costs occur in dependence of the total recorded sample $\nrec$ and the
proportions of the subgroups.
If $\nrec$ is already determined (for some reason) and greater than $n$ from
Equation \eqref{eq_ssc}, the optimal quota $q_0$ can be determined by solving
that equation:
\begin{gather}
    q_0 := 1 \Big/  \! \left\{  \frac{1}{ p_{\mys} \nu_{\mys}^2 }
     \left(  \frac{\nrec \cdot \Delta^2 }{
      (\zalf \kern-1.5pt + \kern-1pt \zbet \kern-1pt )^2 }
      - \nu^2 \right) + 1 \right\}\ \ .
  \end{gather}
In practice, however, $\nrec$ is to be determined and depends on $q_0$. This can
be done numerically by running a simple loop over possible $q$ and checking the
cost function. However, it can also be done analytically: With the model from
Section \ref{sec_MODEL} and using the sample size formula \eqref{eq_ssc} we can
optimise the total costs in relation to $q$:
\begin{align}
    q_0 &:= \operatorname{Opt}(q) = \operatorname{argmin}_q  \! \!
    \left\{ \operatorname{Cost}(n(q),q) \right\}  \\
    &= \operatorname{argmin}_q  \! \!
    \left\{ n(q) \cdot \big( p_\myu \! \! \cdot \!
    c_\myu + p_\mys \! \! \cdot \! \! ( \csn \! + q \! \cdot \! \csz) \! \big)  \right\} \\
    &= \operatorname{argmin}_q  \! \!  \left\{ \!\!
      \Big( \! \zalf \kern-3pt + \kern-2.5pt  \zbet  \kern-1pt  \! \Big)^{\!2} \!\!
      \cdot \!  \frac{    p_{\mys} \kern-1pt \mymu \nu_{\mys}^2  (1 / \kern-1pt q
       \, \text{\texttt{-}}  \, 1 )
     \! + \!  \nu^2 }{ \Delta^2 }
     \! \cdot  \!
     \left( \! \frac{p_\myu c_\myu }{p_\mys \csz} + \frac{ \csn }{\csz}
    + q \! \right) \! \cdot \!  (  p_\mys \csz ) \!\! \right\} \\
    &= \operatorname{argmin}_q  \! \!    \left\{ \!\!
    \frac{ ( \zalf \kern-3pt + \kern-2.5pt  \zbet  \kern-1pt   )^2
     p_\mys \nu_{\mys}^2}{ \Delta^2 } \!\!
      \cdot \! \left( \frac1{q} \, \text{\texttt{-}}  \, \frac{ p_\mys
      \nu_{\mys}^2 }{ p_\mys \nu_{\mys}^2}
       +  \frac{ \nu^2 }{ p_\mys \nu_{\mys}^2}
       \right)       \! \cdot  \!
      \left( q + \! \frac{p_\myu c_\myu }{p_\mys \csz} + \frac{ \csn }{\csz}
     \! \right)  \!\! \right\} \\
     &= \operatorname{argmin}_q  \! \!    \Big\{ \!
      \operatorname{const}
      \cdot  \Big( \frac1{q} + \underbrace{ \frac{ \nu^2  \, \text{\texttt{-}}  \,
      p_\mys \nu_{\mys}^2 }{ p_\mys \nu_{\mys}^2}  }_{=:b} \Big)        \! \cdot  \!
       \Big( q + \! \underbrace{\frac{p_\myu c_\myu }{p_\mys \csz} + \frac{ \csn }{\csz}
     }_{=:a}\! \Big)  \!\! \Big\} \\
\intertext{The derivative of the function $f(q) = (q+a) \! \cdot \! (1/q + b) = b \cdot q
    +  a/q  +  \operatorname{const} \; $ to the variable $q$ results in
   $f'(q) = b + a  \cdot \kern-3pt (-1/q^2)$
   and with $f'(q_{\text{min}}) \overset{!}{=} 0$ it follows, that for all
   $q \geq 0\%$ and $q \leq 100\%$ all non-marginal minima must hold $q_{\text{min}}
   = +\sqrt{a/b} $. Since costs diverge to infinity for $q$ towards 0, the minimum of $f(q)$
   is $\operatorname{min} \! \! \big( \! \sqrt{a/b} , 100\% \! \big)$ and it follows}
    q_0 & = \operatorname{min} \! \left(
      \sqrt{ \left( \frac{p_\myu c_\myu }{p_\mys \csz} + \frac{ \csn }{\csz} \right)
      /   \left( \frac{ \nu^2  \, \text{\texttt{-}}  \,
      p_\mys \nu_{\mys}^2 }{ p_\mys \nu_{\mys}^2}  \right) } , \, 100\% \right)\ \ .
\end{align}
\section{Application}\label{sec_application}
The equivalence test as currently in use is summarized in Procedure
\ref{method_equivalence_test} and our new partitioned equivalence test in
Procedure \ref{method_partitioned_equivalence_test}.

\subsection{Use Cases}\label{sec_use_cases} 
\begin{figure}
  \setlength{\unitlength}{1cm}
  \includegraphics[page=1,width=0.95\textwidth]{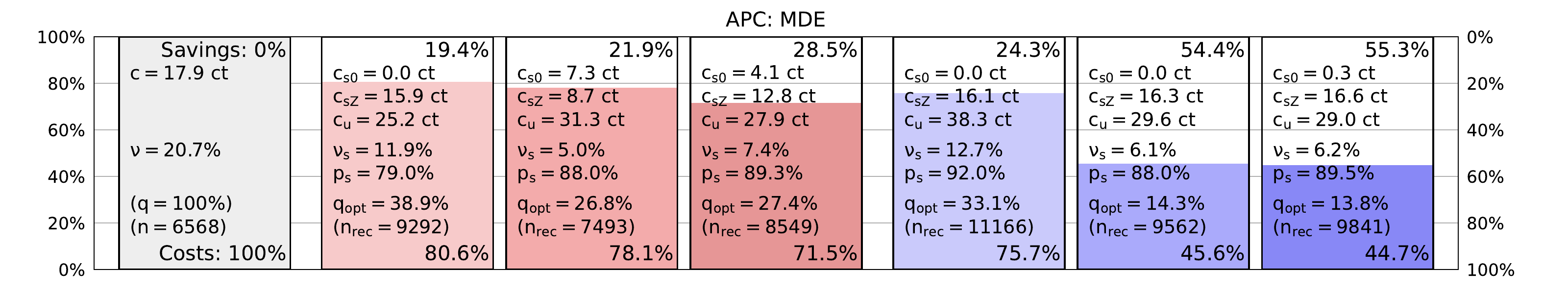}
  \includegraphics[page=1,width=0.95\textwidth]{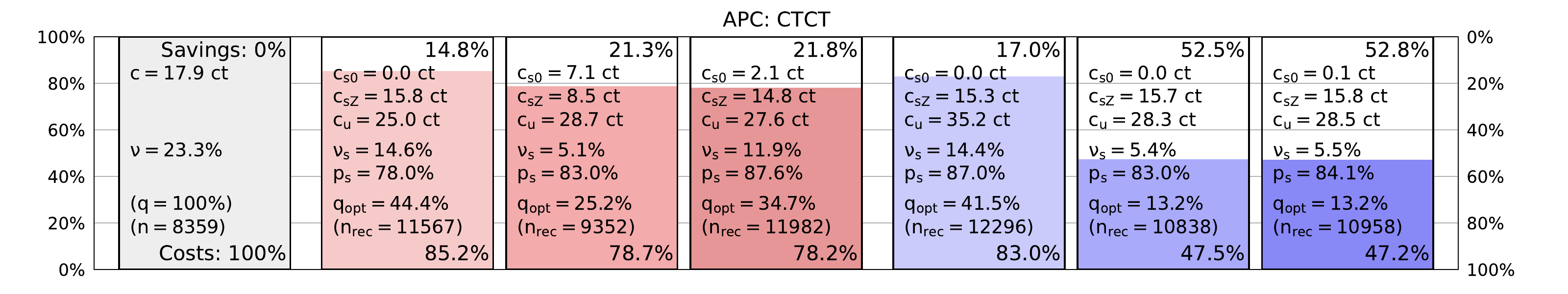}
  \includegraphics[page=1,width=0.95\textwidth]{%
  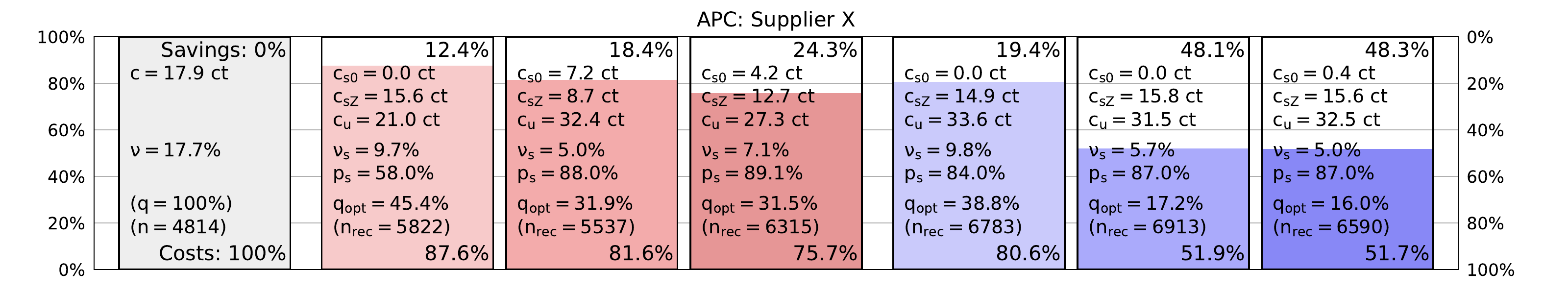}
  \includegraphics[page=1,width=0.95\textwidth]{%
  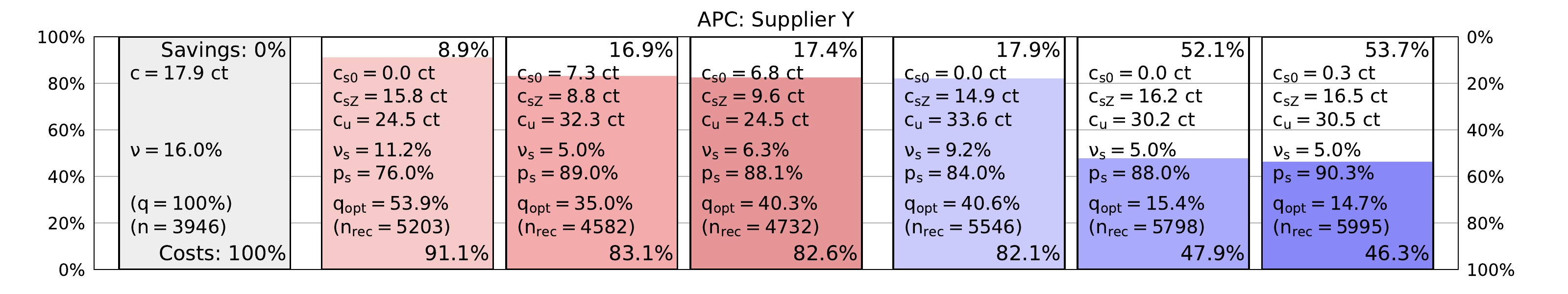}
  \includegraphics[page=1,width=0.95\textwidth]{%
  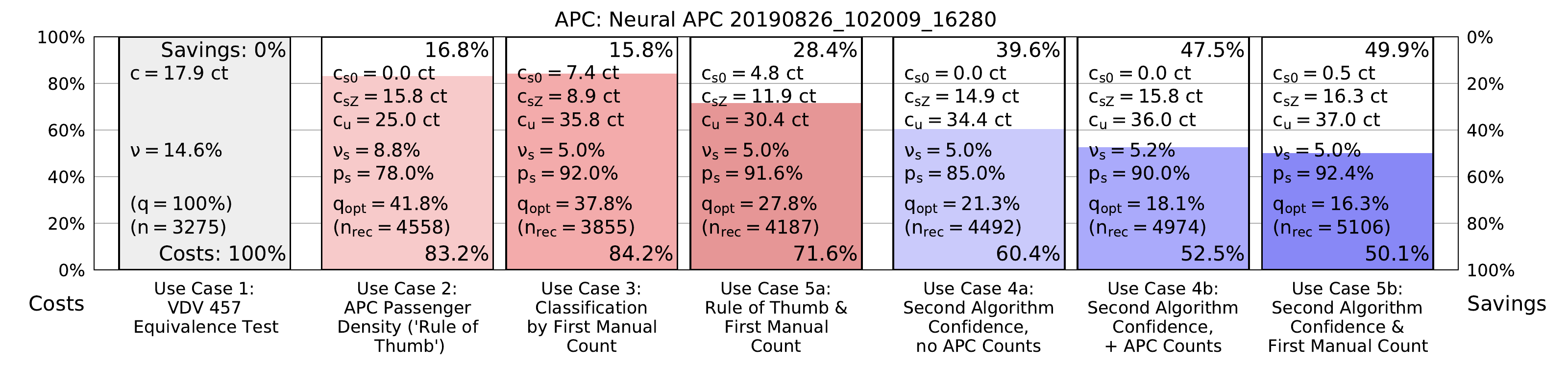}
  \centering
  \caption{Possible savings for different use cases and APC systems. Red bars
  indicate an entirely manual workflow with only sensor counts necessary
  (e.g.\@{} from the APC system vendor) while blue bars indicate that the videos
  from the door opening phases were processed by an additional (possibly non APC
  system vendor) algorithm which yields both additional counts as well as a
  self-estimated confidence of correctness. In our case, this is an ensemble of
  9 Neural APC \citep{Jahn2019} agents. For the manual workflows, the savings
  are between 15\% and 25\% with the first manual count (use cases 3 and 5a)
  being slightly more effective than the rule of thumb (use case 2) solely.
  Still, costs can already be significantly reduced by reorganizing the current
  validation workflow and using the partitioned equivalence test. Using only a
  second algorithm (use case 4a) and interpreting its confidence estimation as a
  general scene difficulty shows mixed results: when the algorithm and the
  estimator share a common background, which is the case for the Neural APC
  20190826\_102009\_16280 and the ensemble of the remaining agents, the savings
  are around 35\%, while otherwise, they are typically below 20\%, which is a
  little less than purely manual obtainable savings (red bars). However, when
  combining the sensor counts with the second algorithm and confidence, around
  50\% savings can be reached in average (use case 4b and use case 5b). For the
  simulation, passenger counts on a combined (2D and 3D) dataset of 11530 videos
  have been used, with at least three manual counts per video (six-eyes
  principle). The costs are stated in euro cent, compare Appendix
  \ref{sec_CostsExamples}.}\label{fig_pe_performance}
\end{figure}

Since the classification function
itself can be completely arbitrary and due to the large amount of possible
partitions (e.g. in the case of a recommended sample size of 6147 there are
$2^{6141}=2.7\cdot 10^{1850}$ possible partitions), we introduce \emph{use
cases}:
\begin{enumerate}
  \item The original equivalence test: either all videos are classified as safe
  or all videos are classified as unsafe. In both cases, the partitioned
  equivalence test reduces to the original equivalence test.
  \item A simple classification, a so-called \emph{rule of thumb}: this use case
  can e.g.\@{} depend on the automatic counts only, like sorting videos by their
  passengers per minute count and creating the partitions according to whether
  a certain threshold has been surpassed or not. By this method, door opening
  phases more prone to overcrowding are considered to be less safe.
  \item Classification by using the first manual count: since in VDV 457 two
  manual counts plus a supervisor as a tie-breaker is required, the first manual
  count can already be used to classify whether a video is safe. Like in the
  second use case, videos are sorted according to their difference in manual and
  automatic count and split into safe and unsafe partitions using a certain
  threshold.
  \item Classification by algorithms/artificial intelligence: the videos are
  processed by an additional algorithm, which produces an estimate of how
  difficult or unsafe it considers the video to be, which are handled like in
  the use cases before.
  \begin{enumerate}
    \item Only use the safeness estimate and ignore the count of the APC system
    entirely. We take a look at this use case to determine whether there is an
    \emph{algorithm independent intrinsic video or scene difficulty}. In case
    the additional algorithm and the APC system's algorithm are related, we
    expect greater savings.
    \item The additional algorithm is capable to create a count as well. Use the
    difference to the APC system's count and the safeness estimate as a
    tie-breaker if the count delta is zero.
  \end{enumerate}
  \item Combined Classification: the methods above can be combined to yield
  better savings than the individual use cases themselves.
  \begin{enumerate}
    \item Use cases 2 and 3: first, classify by a rule of thumb, then do a first
    manual count according to that classification. This approach is still
    entirely manual.
    \item Use cases 3 and 4b: as the case before, but with an algorithm instead
    of a rule of thumb. This approach has higher requirements, but may yield
    greater savings as well.
  \end{enumerate}
\end{enumerate}
For an evaluation of the use cases performance, see Figure
\ref{fig_pe_performance}.

\begin{algorithm}
  \caption{Current Equivalence Test w.r.t.\@{} VDV 457 v2.1}
  \label{method_equivalence_test}
  \begin{enumerate}
    \item \textbf{Parameter specification}:
    $\alpha=\beta=5\%$, $\Delta=1\%$, $\nu=20\%$ \\
    $\nu$ is APC system manufacturer dependent, modern systems can achieve 
    $\nu \le 15\%$
    \item \textbf{Sample size estimation}:
    $n = \nref = (\zalf+\zbet)^2 \frac{\nu^2}{\Delta^2} $
    \item \textbf{Sample size buffer}
    (includes an increase of the sample size by 15\%)
    \item \textbf{Perform the actual, manual comparison count}
    \item \textbf{Evaluation of the
    }$(1\!-\!\alpha)(=95$\%$)$--\textbf{confidence interval } $\left[ \olD \pm
    z_{1-\alpha/2} \cdot \frac{\hnu}{\sqrt{n}} \right]$
    \item \textbf{Check}, if confidence interval is contained entirely within
    $[- \Delta, + \Delta]$:
    \begin{enumerate}[label=(\roman*)]
      \item if yes: equivalence test successfully passed
      \item if no: possibly increase sample size and reevaluate equivalence test
      or consider equivalence test as failed
    \end{enumerate}
  \end{enumerate}
\end{algorithm}
\begin{algorithm}
  \caption{Partitioned Equivalence Test}
  \label{method_partitioned_equivalence_test}
  \begin{enumerate}
    \item \textbf{Test parameter specification}:
    $\alpha=\beta=5\%$, $\Delta=1\%$, $\nu=15\%$ \\
    $\nu$ is APC system manufacturer dependent, modern systems can achieve
    $\nu \le 15\%$ \\
    To protect user risk $\le\alpha$ against inappropriately low sample sizes,
    e.g.\@{} $\numin=3\%$ can be introduced
    \item \textbf{Partition parameter specification}:
    $0\le p_{\mys}\le 1$, $\nu_{\mys}/\nu$ and $0 \le q \le 1$,\\
    e.g.\@{} $p_{\mys}=90\%$, $\nu_{\mys}/\nu=35\%$, $q=17.5\%$ for
    algorithmic and $q=35\%$ for purely manual use cases

    (Actual values can be specified by the classification method provider. \\
    User risk $\le \alpha$ is guaranteed like in the non-partitioned equivalence
    test.)
    \item \textbf{Sample- resp.\@{} record size estimation}:
    \begin{align*}n =\nrec = \nref \kern-1pt \cdot \biggl[ p_{\mys}
      \frac{\nu_{\mys}^2}{\nu^2} \Bigl(\frac{1}{q}-1\Bigr)+1\biggr] \qquad
      \text{with} \qquad \nref = (\zalf+\zbet)^2 \frac{\nu^2}{\Delta^2} \end{align*}
    \item \textbf{Sample size buffer}
    (includes an increase of the sample size by 15\%)
    \item \textbf{Record the counting material} (e.g.\@{} 3D-videos)
    \item \textbf{Use classification method to partition the counting material}
    \item \textbf{Perform the manual count w.r.t.\@{} the partition}

    Count all $N_{\myu}$ door opening phases in the unsafe partition

    Count $N_{\mys}\cdot q$ random door stop
    events in the safe partition
    \item \textbf{Evaluation of the
    }$(1\!-\!\alpha)(=95$\%$)$--\textbf{confidence interval} $\left[ \olD \pm
    z_{1-\alpha/2} \cdot \frac{\hnu}{\sqrt{n}} \right]$ \textbf{with}
    \begin{align*} \textstyle
      \hM_q &= \hp_\mys \mymu \hM_{\mys q} + \hp_\myu \mymu \olM_\myu\ \ ,
      \qquad D_i := (K_i - M_i) / \hM_q\quad \text{für}\quad \olD_{\mys q},
      \ \olD_\myu,\ \hnu_{\mys q},\ \hnu_{\myu}\\
      \olD &= \hD_q = \frac{N_{\mys}}{n} \mymu \olD_{\mys q}
      + \frac{N_{\myu}}{n} \mymu \olD_\myu  \\
      \hnu^2 &= \hnu_q^2 = \frac{N_{\mys} }{n} \mymu \frac{\max(\hnu_{\mys q},
      \numin)^2}{q}
      + \frac{N_{\myu} }{n} \mymu \max(\hnu_{\myu}, \numin)^2 +  \frac{N_{\mys}
      \mymu N_{\myu} }{n^2} \mymu \bigl(\olD_{\mys q} - \olD_{\myu}\bigr)^2
    \end{align*}
    \item \textbf{Check}, if confidence interval is contained entirely within
    $[- \Delta, + \Delta]$:
    \begin{enumerate}[label=(\roman*)]
      \item if yes: partitioned equivalence test successfully passed
      \item if no: possibly increase sample size and reevaluate partitioned
      equivalence test or consider partitioned equivalence test as failed
    \end{enumerate}
  \end{enumerate}
\end{algorithm}
\subsection{Discussion and Suggested Values}
As can be seen in Figure \ref{fig_pe_performance}, basically two use cases
remain: the entirely manual use case 5a and use case 4b. Use case 2, the rule of
thumb, is not effective enough and use case 3, the first manual count can often
significantly profit from a pre-classification, turning it into use case 5a. For
the algorithmic partitioned equivalence test, combined cases do not yield a lot
of improvement, yet complicating the process.

Overall, as optimal parameters, $p_s=90\%$ is a common finding and
$\nu_s/\nu=35\%$ can be assumed. For $q$, in purely manual methods, $q=35\%$
seems suiteable, while $q=17.5\%$ proved viable for the algorithmically assisted
use cases (compare figures in Appendix \ref{sec_VarianceSmall}).
\section{Conclusion}\label{sec_conclusion} Our investigations have shown that a
seamless connection of the partitioned equivalence test to current requirements
from \vdvafzs{} (and the equivalence test itsself) is possible. Here the
standard benefits considerably from the completed changeover from the t-test to
the equivalence test. The partitioned equivalence test allows statistically
robust samples to be realized at significantly reduced costs when compared to
the classic equivalence test. Not only can costs be reduced, but also more
manual tests can be carried out with the same monetary investment, thus
increasing quality. Measured by the volume of funds to be distributed by revenue
sharing, shortcomings in passenger counting validation are not justifiable in
economic terms anyway. The partitioned equivalence test is not exclusively
limited to the use of algorithms, as shown by the classification by the first
manual counting, a rule of thumb or a combination of both. Basically, any
information available about the counting behavior of the sensors can be used to
reduce costs -- in a statistically robust way and thus without increased risk
for the user. There is also no vendor lock-in, as all relevant calculations can
be performed with simple spreadsheet formulas without specialized software. In
the medium to long term, the partitioned equivalence test lays the foundation
for a deeper integration between manual and automatic counting to raise count
quality to a new level. This will not only benefit revenue distribution, but
also forecasting and real-time in-vehicle passenger counts. The new test will be
an incentive for transport companies to invest in the corresponding IT
infrastructure. In other industries, e.g.\@{} the tech industry, hybrid systems,
in which people and algorithms work together, are already common, e.g.\@{} in
fraud detection. We hope that the partitioned equivalence test will not only
help automatic passenger counting catch up with state-of-the-art technologies,
but will even make it a technological pioneer other fields can profit from.
\section*{Acknowledgements}
  This research is financially supported by the European Regional Development
  Fund.
\section*{Authors' contribution}
D Ellenberger: Statistics Lead, Partitioned Equivalence Test Formalization \&
Formal Proofs, Literature Search and Review, Illustration and Code Prototypes,
Data Processing \& Analysis, Manuscript Writing and Editing.\\
M Siebert: Research Lead, Partitioned Equivalence Test Concept, Illustrations
and Code, Data Processing \& Analysis, Simulations, Creation of VisualCount \&
Depth Sensing Unit, Manuscript Writing and Editing.
\section*{Conflict of Interest}
David Ellenberger has been employed by Interautomation Deutschland GmbH during
the time of research and manuscript preparation. None resulted in a conflict of
interest. Michael Siebert is an employee of Interautomation Deutschland GmbH.
The submitted work does not pose a conflict of interest.
%
\bibliography{bibliography/bibliographySTAT,bibliography/bibliographyMIX,bibliography/bibliographyAPC,bibliography/bibliographyUSAGE}
\bibliographystyle{spbasic}

\newpage
\begin{appendix}

  \section{Confidence intervals for sums with random weights}\label{sec_KI}
  In the following we use the notation of Section 2.2. Let $ \mathds{X} =
  (X_i)_{i=1,..,n}$ the counting errors of an APC system, independent, identically
  distributed with probability measure $\myp^X$, $\E(X_1) = \mu$ and $\Var(X_1) =
  \nu$. The indicator $W_i$ represents the (random) classification into safe and
  unsafe door opening phases (DOP). Then, $W_i$ is Bernoulli($p_s$)-distributed, in
  the sense that the outcome is \emph{safe} with likelihood $p_s$ and
  \emph{unsafe} with likelihood $p_\myu \myeq \myom  p_\mys$ ($W_i \sim \myp^W
  \sim \operatorname{Bin}(1, \kern1pt p_s)$). The total number of safe DOP is
  $N_\mys = \sum W_i \left( = \sum \mathds{1}_{\left\{\kern-1pt  W_i  \myeq \mys
  \kern-1pt \right\} } \right) $ which is the sum of all $W_i$ and $\hp_{\mys} =
  N_\mys / n$. We now consider the random variables $X_i \myv W_i$, i.e. the
  distribution of errors in the case that a DOP is safe resp. unsafe. The
  associated probability measure of the conditional distribution $X \myv W$ always
  exists, since $W$ is integer and further the assumption $\myp(W_i \myeq \mys)
  \kern-3pt > \kern-3pt 0 $ holds, since otherwise it would be trivial.
  Note that $X_i$ and $W_i$ are not needed to be stochastically independent. Let
  $\E(X_i \, \myv W_i \myeq \myu ) = \mu_\myu$ and $\E(X_i \, \myv W_i \myeq \mys)
  = \mu_\mys$. Then for the expected value the following holds:
   \begin{align}
    \E( X_1 ) &= \int \kern-9pt \int x \; \mydp^{X \myv  W } \! \!
      ( \kern-1pt x \kern-1pt ) \,  \mydp^W \\
    &= P(W_1 \myeq \mys  )\mymu \! \!  \int x \, \mydp^{X \myv W \! \myeq \mys } \,
      +\, \myp(W_1 \myeq  \myu ) \mymu \!\! \int x \, \mydp^{X \myv W \! \myeq \myu }\\
    &= p_\mys \mymu \mu_\mys  + p_\myu \mymu  \mu_\myu
  \end{align}
  Further, let now be $i_{\mys}(k)$ for $k \myeq 1,..,N_\mys$ the assignment in
  ascending order to the (random) variables of the safe DOP, such that for
  $i=i_{\mys}(k)$ the notation $X_{\mys k} := X_i$, $W_{\mys k} := W_i$, etc. is
  well defined and $i_{\myu}(k)$ for $k \myeq 1,..,N_\myu$ the analogous
  assignment to the variables of the unsafe DOP ($X_{\myu k}$,$W_{\myu k}$, etc.).
  Methods are now required to construct a range-preserving confidence interval if
  only a previously specified proportion $q_0$ (e.g. $50 \%$) of the safe DOP to
  be manually counted. First an adjustment (upwards) of the pre-specified $q_0$
  takes place, such that the number of safe DOP to be counted is integer $q :=
  \lceil q_0 \mymu N_\mys \rceil / N_\mys $.
  
  A random sample of $q \mymu N_\mys$ safe DOP can be implemented with $\mathds{Z}
  \myeq (Z_k)_{1,..,N_\mys} \! \! \in \! \left\{0,1 \right\}$, such that
  $\sum_{k=1}^{N_\mys} Z_k = q \mymu N_\mys$ holds. The resulting
  $(Z_k)_{1,..,N_\mys}$ are by design independent of all measured variables. For
  $\iton$ let with $i = i_{\mys}(k)$ the weights be defined $C_i = Z_k \kern-1pt /
  \kern-1pt q $ for the safe DOP and with $i = i_{\myu}(k)$ the corresponding $C_i
  = 1$ for the unsafe DOP.
  
  The estimator for $\mu_\myu$ will be the mean value
  \begin{align}
   & \overline{X}_\myu  = \frac1{N_\myu} \sum_{k=1}^{N_\myu} X_{\myu k}
   \intertext{and for $\mu_\mys$ one can state a similar estimator,
   which we will refer to as leave $(\myom q)$ out estimator:}
   & \hX_{\mys q} = \frac1{N_\mys} \sum_{k=1}^{N_\mys} X_{\mys k}
      \cdot  Z_k  \kern-1pt / \kern-1pt q
   \intertext{An estimator for $\mu$ can now be constructed from the individual 
   estimators as follows:}
   & \hX_q = \hp_\mys \mymu \hX_{\mys q} + \hp_\myu \mymu \olX_\myu
  \end{align}
  All values $X_{\myu k}$ where $Z_k$ is zero are not needed. These are the values
  that are not needed for the calculation, i.e. can be discarded. Then,
  \begin{align}
   \E (\hX_q) &= \frac1n \cdot \E\left( \frac{N_\mys}{N_\mys}
   \sum_{k=1}^{N_\mys} X_{\mys k} \cdot  Z_k  \kern-1pt / \kern-1pt q
   \, + \, \frac{N_\myu}{N_\myu} \sum_{k=1}^{N_\myu} X_{\myu k} \right) \\
   &= \frac1n \cdot \left( p_{\mys} \kern-1pt \mymu n \cdot \mu_{\mys} \!  \mymu \E(Z_k)
        \kern-1pt / \kern-1pt q   + p_{\myu} \kern-1pt \mymu n  \cdot \mu_{\myu} \right) \\
   &= p_{\mys} \mymu \mu_{\mys} + p_{\myu} \mymu \mu_{\myu} = \mu
  \end{align}
  and thus $\hX_q$ is an unbiased estimator which can be calculated on a reduced
  data set. For the variance of the estimator the following holds:
  \begin{align*}
       \Var (\hX_q)
       &= \E(\hX_q \! - \mu)^2    = \int \kern-9pt \int (\hX_q \! - \mu)^2
            \, \mydp^{ \mathds{X} \, \myv  W } \! \!  \,  \mydp^W \\
       &= \frac1{n^2} \int \kern-9pt \int \left( N_{\mys} \kern-1pt \mymu \hX_{\mys q} +
             N_{\myu} \kern-1pt \mymu \olX_{\myu} \; - n \mymu \mu \right)^{\!\!2}
             \, \mydp^{ \mathds{X} \, \myv  W } \! \!  \,  \mydp^W \\
       &= \frac1{n^2} \int \kern-9pt \int \left(\sum_{k=1}^{N_\mys}
            ( X_{\mys k} \cdot Z_k  \kern-1pt / \kern-1pt q - \mu ) +
            \sum_{k'=1}^{N_\myu} (X_{\myu k'}  - \mu ) \right)^{\!\!\!2}
            \, \mydp^{ \mathds{X} \, \myv  W } \! \!  \,  \mydp^W \\
       &= \frac1{n^2} \! \int \kern-9pt \int \! \left(\sum_{k=1}^{N_\mys}
           ( X_{\mys k} \mymu Z_k  \kern-1pt / \kern-1pt q   \text{\texttt{-}}
           \kern1pt  \mu_{\mys} \text{\texttt{+}} (p_{\myu} \! \mymu \mu_{\mys}
            \kern-1pt  \text{\texttt{-}} \kern1pt  p_{\myu} \!  \mymu \mu_{\myu} ) )
            + \sum_{k'=1}^{N_\myu} (X_{\myu k'}    \text{\texttt{-}} \kern1pt
            \mu_{\myu} \text{\texttt{+}} p_{\mys} \! \mymu (\mu_{\myu}  \kern-1pt
             \text{\texttt{-}} \kern1pt  \mu_{\mys} ) ) \!  \right)^{\!\!\!2} \!
             \mydp^{ \mathds{X} \, \myv  W } \! \!  \,  \mydp^W
      \intertext{with $\delta := \mu_{\mys} \kern1pt \text{\texttt{-}} \kern1pt
          \mu_{\myu}$ und $ \sum_{k=1}^{N_\mys} Z_{k} \kern-1pt / \kern-1pt q \mymu
          \mu_{\mys} = N_\mys \mymu \mu_{\mys} $ it follows:}
       &= \frac1{n^2} \! \int \kern-9pt \int \! \left(\sum_{k=1}^{N_\mys} Z_k
          \kern-1pt / \kern-1pt q  \mymu ( X_{\mys k}  \text{\texttt{-}} \kern1pt
          \mu_{\mys}  ) \!  + \!  \sum_{k'=1}^{N_\myu} (X_{\myu k'}  \text{\texttt{-}}
          \kern1pt  \mu_{\myu}  )  +  N_\mys \! \mymu  p_{\myu}
          \mymu \delta \kern1pt \text{\texttt{-}} \kern1pt   N_\myu \! \mymu
          p_{\mys}  \mymu \delta    \right)^{\!\!\!2}
          \mydp^{ \mathds{X} \, \myv  W } \! \!  \,  \mydp^W \\
       \intertext{with $N_\mys \! \mymu  p_{\myu}  \mymu \delta \kern1pt
            \text{\texttt{-}} \kern1pt   N_\myu \! \mymu  p_{\mys}  \mymu \delta
            = (N_\mys \mymu ( \myom  p_{\mys} \kern-1pt )  \kern-0pt
            \text{\texttt{-}} \kern1pt ( n \kern+1pt  \text{\texttt{-}}
            \kern1pt N_\mys \kern-1pt ) \mymu  p_{\mys}  )\mymu \delta
            = (N_\mys \text{\texttt{-}} \kern1pt n \mymu  p_{\mys} ) \mymu \delta$
            and union of the sums of the safe and unsafe DOP it follows:}
       &= \frac1{n^2} \! \int \kern-9pt \int \! \Biggl(\sum_{i=1}^{n}  C_i \mymu
            \Bigl( X_i \, \myv W_i - \underbrace{ \E( X_i\, \myv W_i) }_{=: \,
            \mu_i^{\mys \! / \!  \myu} } \Bigl) \quad  +  \quad
            ( N_\mys  \kern0pt \text{\texttt{-}} \kern1pt  n \mymu  p_{\mys}  )
            \mymu \delta    \Biggl)^{\!\!\!2}
            \mydp^{ \mathds{X} \, \myv  W } \! \!  \,  \mydp^W \\
       &= \frac1{n^2} \! \int \kern-9pt \int \! \Bigl( \underbrace{ \sum_{i=1}^{n}
            C_i \mymu ( X_i \, \myv W_i - \mu_i^{\mys \! / \! \myu} ) }_{=:\, A }\Bigl)^{\!\!2}
            \quad + \quad \underbrace{ \vphantom{\sum_{1}^{n} } 2 \mymu A \mymu
            B}_{\rightarrow 0} \quad + \quad  \Bigl( \underbrace{ \vphantom{ \sum_{1}^{n} }
            \! ( N_\mys  \kern0pt \text{\texttt{-}} \kern1pt  n \mymu  p_{\mys}  )
            \mymu \delta \! }_{=: \, B}   \Bigl)^{\!\!2}
            \mydp^{ \mathds{X} \, \myv  W } \! \!  \,  \mydp^W \\
       &= \frac1{n^2} \! \int \kern-9pt \int \!  \sum_{i=1}^{n} \bigl[ C_i
            \mymu \bigl( X_i \, \myv W_i -  \mu_i^{\mys \! / \!  \myu}  \bigl) \bigl]^2
            + \sum_{i \neq i'}^{n}  C_i \mymu   C_{i'}  \mymu
            \underbrace{ \bigl( X_i   \myv W_i \! - \! \mu_i^{\mys \! / \!  \myu}  \bigl)
            \mymu  \bigl( X_{i'}   \myv W_{i'} \! - \! \mu_{i'}^{\mys \! / \!  \myu}
             \bigl) }_{ \rightarrow 0}    +B^2 \,
             \mydp^{ \mathds{X} \, \myv  W } \! \!    \mydp^W \\
       &=  \frac1{n^2} \left( \int  n \mymu  \Var \bigl( C_1 \mymu ( X_1 \, \myv W \!
            - \! \mu_i^{\mys \! / \!  \myu}  ) \bigl) \mydp^W + \delta^2 \! \mymu \! \!
            \int ( N_\mys  \kern0pt \text{\texttt{-}} \kern1pt  n \mymu  p_{\mys} )^2
            \mydp^W \right) \\
       &= \frac1{n} \; \;   \Bigl( \; \Var \bigl( Z_1 \kern-1pt / \kern-1pt q \, \mymu
            ( X_{\mys 1}  \text{\texttt{-}} \kern1pt   \mu_{\mys}  ) \bigr) \mymu p_{\mys}
            + \Var\bigl( X_{\myu 1}  \kern0pt \text{\texttt{-}} \kern1pt \mu_{\myu} \bigr)
            \mymu p_{\myu} \quad + \quad \delta^2 \mymu p_{\mys} \! \mymu p_{\myu} \Bigr) \\
      \end{align*}
      Furthermore, for the variance of a product of independent random variables
   $C$ and $X$ the identity $\Var(C \mymu X) = \E(C)^2 \mymu \Var(X) +  \E(X)^2
   \mymu \Var(C) + \Var(C) \mymu \Var(X)$ holds and thus
      \begin{align*}
        \Var(Z_1 \kern-1pt / \kern-1pt q \mymu  ( X_{\mys 1} \text{\texttt{-}} \kern1pt \mu_{\mys} ) )
        &= \underbrace{ \E( Z_1 \kern-1pt / \kern-1pt q )}_{=  1} \kern-2pt^2  \mymu \Var( X_{\mys 1} )
        \! + \! \underbrace{\E( X_{\mys 1}  \kern0pt \text{\texttt{-}} \mu_{\mys} )}_{=  0} \kern-2pt^2 \mymu
          \Var(Z_1 \kern-1pt / \kern-1pt q) \! + \! \Var(Z_1 \kern-1pt / \kern-1pt q) \mymu \Var( X_{\mys 1})\\
        &= \Var( X_{\mys 1} ) \kern-1pt / \kern-1pt q  \quad  \text{,}
      \end{align*}
      because $Z_1$ has a hypergeometric distribution and thus $\Var(Z_1 \kern-1pt
      / \kern-1pt q) = 1 \kern-1pt / \kern-1pt q  \kern1pt \text{\texttt{-}} 1$
      holds. Thus,
      \begin{align}
       \Var (\hX_q) &= \frac1{n}   \Bigl( \,  p_{\mys} \kern-1pt \mymu \kern-1pt
       \Var \bigl(   X_{\mys 1} \bigr) \kern-1pt / \kern-1pt q   \,
       + (\myom p_{\mys}) \mymu \Var\bigl( X_{\myu 1} \bigr)  \;
       + \;  (\mu_{\mys} \kern1pt \text{\texttt{-}} \kern1pt \mu_{\myu}  \kern-1pt )^2
       \kern-1pt \mymu p_{\mys} \kern-2pt \mymu \kern-1pt (  \kern-1pt \myom p_{\mys})  \Bigr)
      \end{align}.

\section{Practical guidance on parameters for cost functions}
\label{sec_CostsExamples}
For calculations regarding the planning of an APC validation, a plausible cost
function must be assumed:
\begin{enumerate}[label=(\alph*)]
\item Video recording time $t_{ \{\text{video}, \, i\}}$ in hours
\item The acceleration factor $\rAV=t_{\text{labor}}/t_{\text{video}}$ of the
video corresponds to the ratio of working time $t_{\text{labor}}$ to recording
time $t_{\text{video}}$, i.e.\@{} $\rAV<1$  corresponds to an acceleration,
$\rAV=1$, that viewing the footage takes just as long as its duration 
and $\rAV>1$ corresponds to a slowdown compared to the video duration.
\item The labor cost $c_{\text{labor}}$, which is usually equal to the hourly
wage of the manual comparison counters.
\item Additional costs due to the second manual count and the supervisor:
$r_{\text{S}}$
\end{enumerate}
Then the following basic costs are derived from this:
\begin{gather}
  c_{(\text{Z}, i)} := t_{ \{\text{video}, \, i\}} \cdot \rAV  \cdot c_{\text{labor}}
\end{gather}
Empirical results with manual counts, which were performed with untrained
personnel and with software optimized for the workflow, allow the following
approximate values:
\begin{gather}
  \rAV = 0.7 \qquad \qquad  c_{\text{labor}} = 20 \text{\euro}
   \qquad \qquad r_{\text{S}} = 1.2   \,  .
\end{gather}
This results in average counting costs of
\begin{gather}
 \textstyle \frac1n \!\sum_{i=1}^n c_{(\text{Z},i)}=0.164 \,\text{\euro}
\end{gather}
per door opening phase (DOP). If the recording costs are assumed to be zero, the total costs of the
uncertain DOP are calculated as follows:
\begin{align}
   c_{\myu} &= (1 \! + \! r_{\text{S}}) \cdot  \textstyle \frac1{N_\myu} \!
    \sum_{i=1}^{N_\myu}   c_{(\text{Z}, \myu i)}   \ \ ,\\
\intertext{while two cases must be distinguished here for the safe DOP.
 The first case represents the classification without manual counting:}
  \csn &= 0 \label{eq_no_manual_count_first_csn} \\
  \csz &=  (1 \! + \! r_{\text{S}}) \cdot  \textstyle \frac1{N_\mys} \!
  \sum_{i=1}^{N_\mys}   c_{(\text{Z}, \mys i)} \label{eq_no_manual_count_first_csz}
\intertext{The second case when the first manual count is included:}
  \csn &=  \textstyle \frac1{N_\mys} \! \sum_{i=1}^{N_\mys} c_{(\text{Z}, \mys i)} \label{eq_manual_count_first_csn}  \\
  \csz &=  r_{\text{S}} \cdot  \textstyle \frac1{N_\mys} \!
  \sum_{i=1}^{N_\mys}   c_{(\text{Z}, \mys i)}    \ \ . \label{eq_manual_count_first_csz}
\end{align}
Cases in which only a partial count is performed are calculated analogously as
partial totals over the respective DOP. For the use cases described in Section
\ref{sec_use_cases}, different classification options result in safe and
unsafe DOP. The better a classification succeeds, with
simultaneously low costs, the greater is the overall savings potential of the
process. A special approach here is the \emph{combined classification}, e.g.\@{}
consisting of cases 2 and 3: a rule of thumb and the first manual count. With
such a combined classification, the costs incurred can also be determined in a
simple manner. This is done by considering a more detailed partitioning: the
\emph{safe DOP} are those that have already been classified as safe using the
first classification rule (e.g.\@{} rule of thumb) and those that were initially
classified as unsafe, but were checked using the second classification rule
(here the manual count) and were reclassified as \emph{safe} due to lack of
discrepancies in the counts. The quantity of reclassified DOP 
\begin{gather}
    \mathbb{W}_{\text{reclass.} , \mys i} =
    \Bigg\{\begin{array}{ll}
      1 & \text{if DOP $i$ is reclassified as safe after the first classification} \\
      0 & \text{if DOP $i$ is initially classified as safe}
      \end{array}
  \end{gather}
is important for appropriate allocation of costs. The counting costs incurred by
the reclassified DOP for the initial manual (comparative) count are now added to
the basic costs, since they may have been incurred without the ground truth for
that DOP actually being finally determined. In contrast, this attribution does
not happen for \emph{unsafe} DOP, where both classification rules have provided
an unsafe classification, since for these the ground truth must be determined in
any case. Since full counting costs are therefore always incurred here, the
sequence is irrelevant. In summary, the costs of the safe DOP are
as follows:
\begin{align}
  \csn &=  \textstyle \frac1{N_\mys} \! \sum_{i=1}^{N_\mys}
        \mathbb{W}_{\text{reclass.} , \mys i}  \cdot c_{(\text{Z}, \mys i)}   \\
  \csz &=   \textstyle \frac1{N_\mys} \!   \sum_{i=1}^{N_\mys}
  (1- \mathbb{W}_{\text{reclass.}, \mys i}  + r_{\text{S}} )  \cdot c_{(\text{Z}, \mys i)}  \ \ .
\end{align}
An application of this approach is shown in Figure \ref{fig_pe_performance}.

\section{Small-sample problems in variance estimation}\label{sec_VarianceSmall}
\begin{figure}
    \setlength{\unitlength}{1cm}
    \includegraphics[page=1,width=0.95\textwidth]{%
        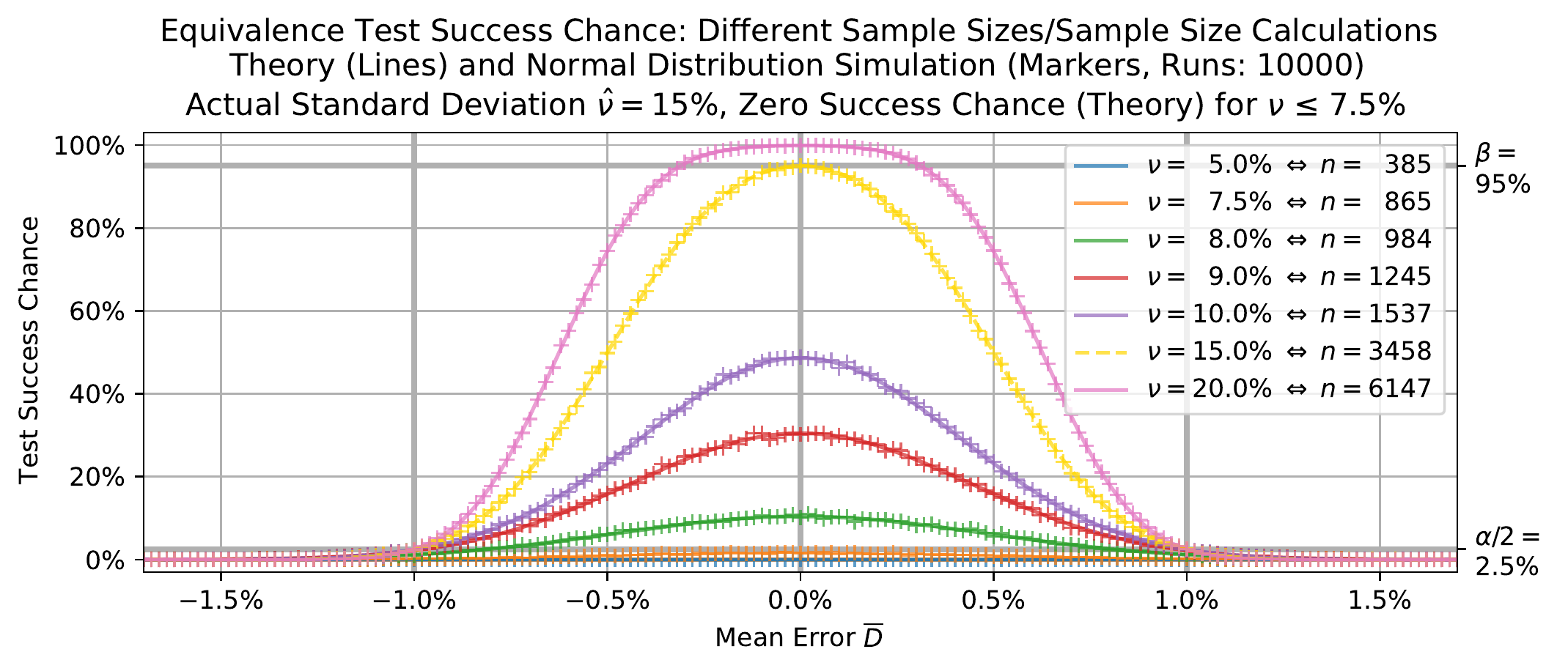}
    \centering
    \caption{Equivalence Test Success Theory and Simulation with a normal
    distribution. The overlap is almost perfect: at $\nu=7.5\%$, the success
    chance is supposed to be zero analytically, from which the simulation
    deviates slightly. For a comparison with actual passenger counting system
    errors, see Figure \ref{fig_test_success_simulation_no_numin}, Figure
    \ref{fig_test_success_simulation_numin_3} and Figure
    \ref{fig_test_success_simulation_numin_5}.}
    \label{fig_test_success_theory}
\end{figure}

\begin{figure}
    \setlength{\unitlength}{1cm}
    \includegraphics[page=1,width=0.95\textwidth]{%
        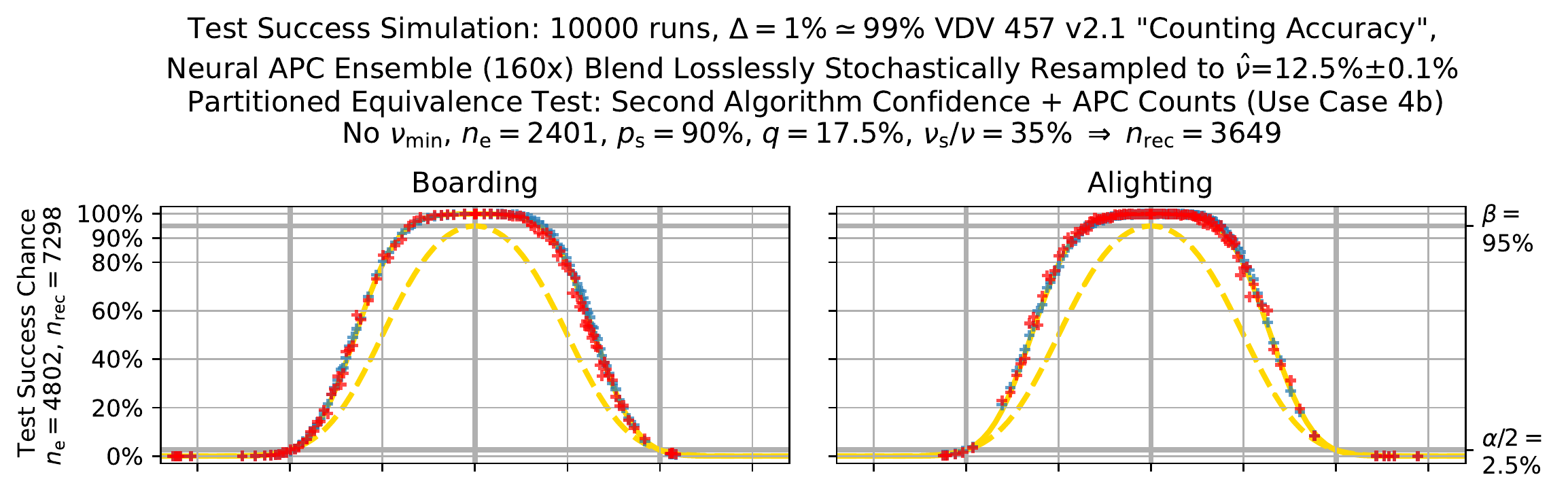}
    \includegraphics[page=1,width=0.95\textwidth]{%
        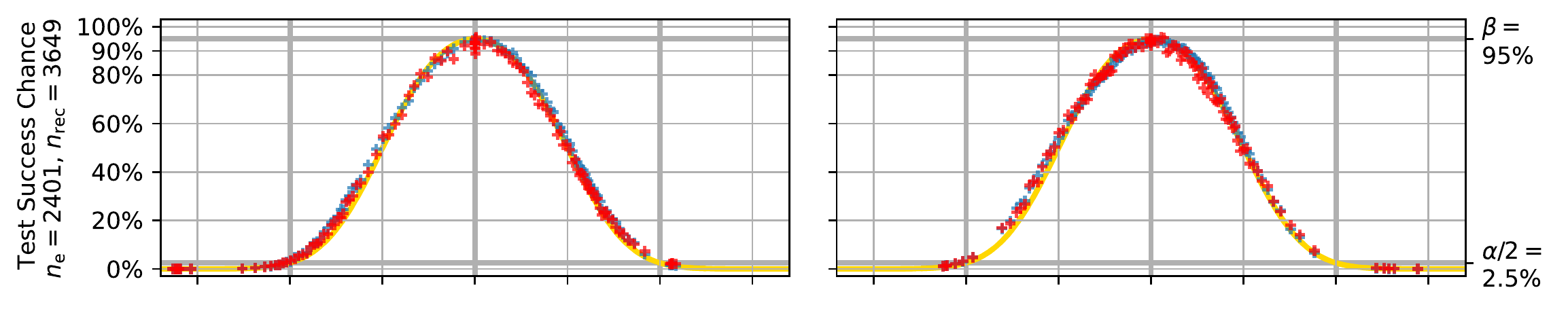}
    \includegraphics[page=1,width=0.95\textwidth]{%
        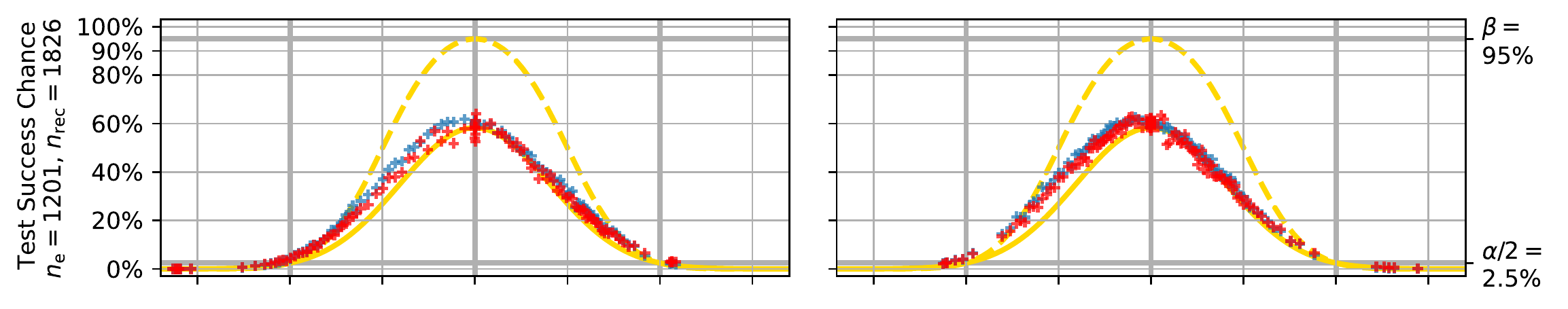}
    \includegraphics[page=1,width=0.95\textwidth]{%
        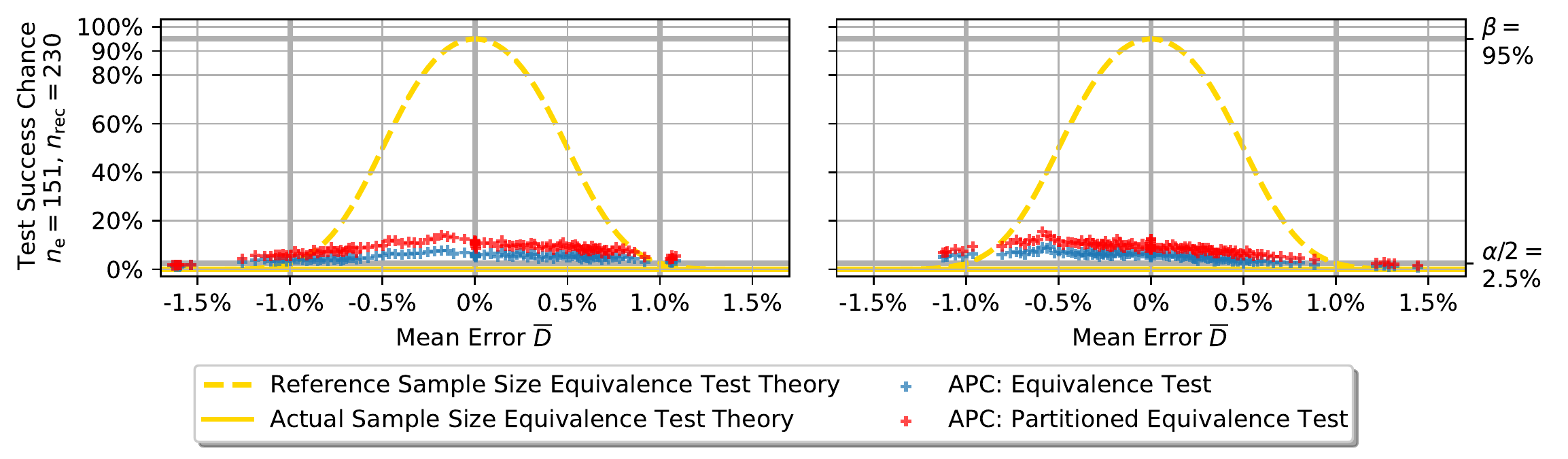}
    \centering
    \caption{Test Success Simulation, comparing the equivalence test with the
    partitioned equivalence test, no $\numin$ resp.\@{} $\numin=0\%$. The
    difference of $\olD$ to a perfect normal distribution (compare Figure
    \ref{fig_test_success_theory}) is that it has a slightly higher chance of
    test success than it should have for low sample sizes. Even though success
    chances of around $10\%$ to $15\%$ across all $\olD$ will not allow any APC
    manufacturer any kind of sustainable business, in order to ensure the
    bounded user risk promise of success chance $\le\alpha/2=2.5\%$ for $|\olD|
    \ge \Delta = 1\%$, the introduction of $\numin>0$ can close the theoretical
    gap as can be seen in Figure \ref{fig_test_success_simulation_numin_3} and
    Figure \ref{fig_test_success_simulation_numin_5}.}
    \label{fig_test_success_simulation_no_numin}
\end{figure}

\begin{figure}
    \setlength{\unitlength}{1cm}
    \includegraphics[page=1,width=0.95\textwidth]{%
        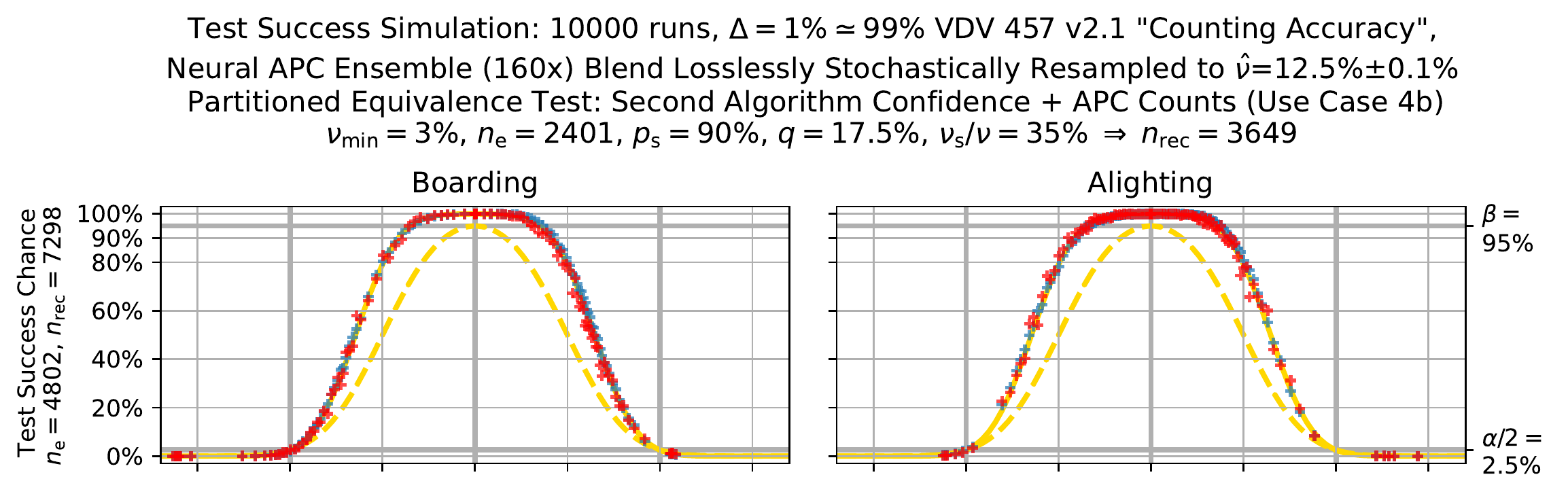}
    \includegraphics[page=1,width=0.95\textwidth]{%
        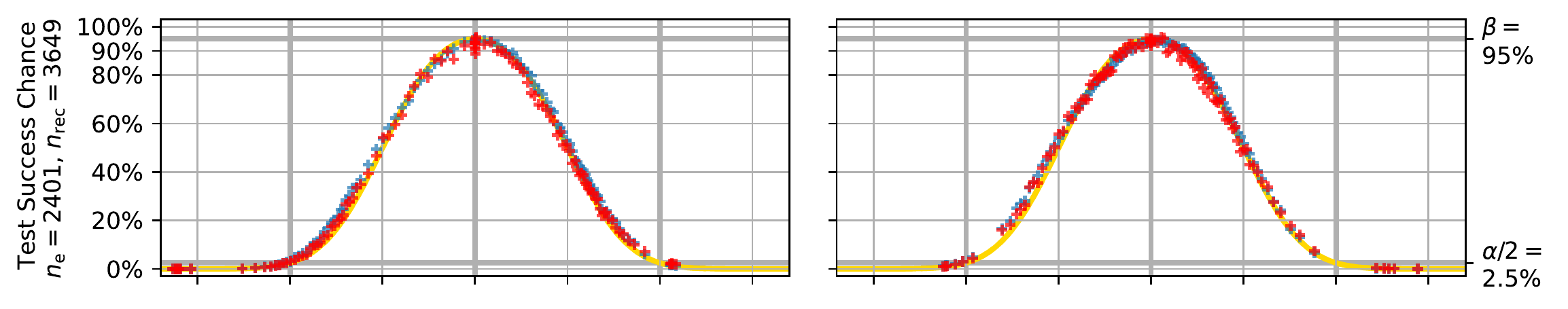}
    \includegraphics[page=1,width=0.95\textwidth]{%
        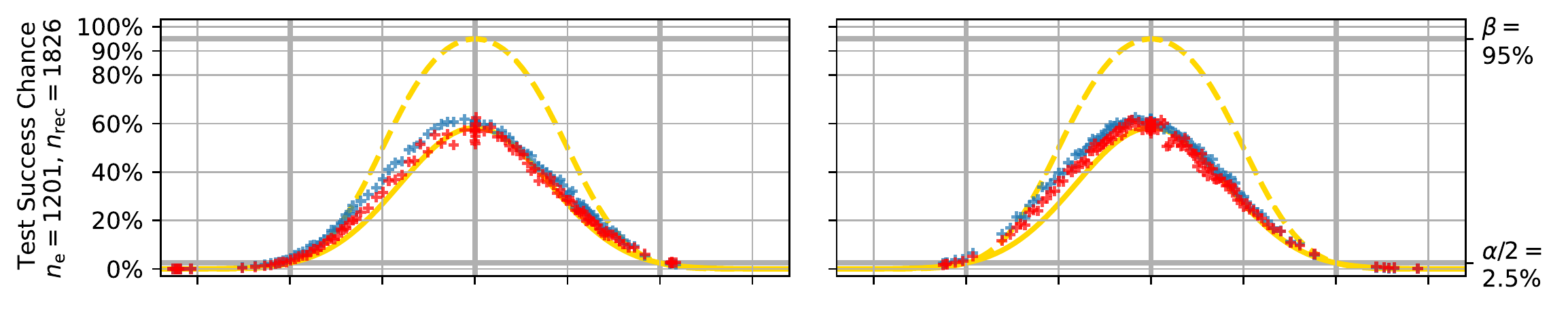}
    \includegraphics[page=1,width=0.95\textwidth]{%
        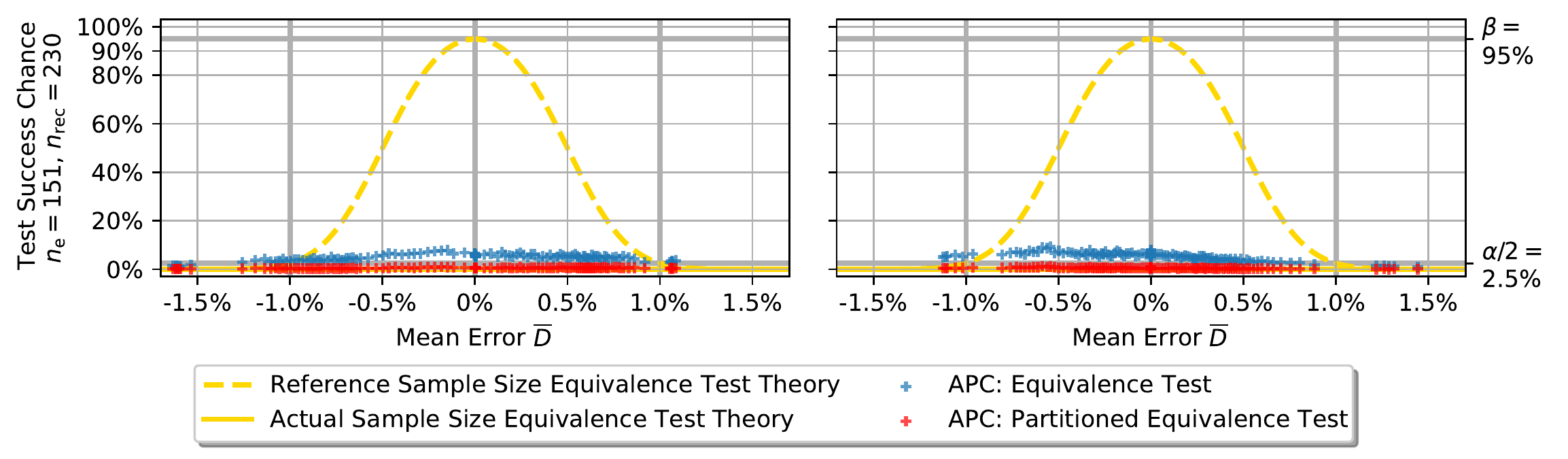}
    \centering
    \caption{Test Success Simulation, comparing the equivalence test with the
    partitioned equivalence test, $\numin=3\%$. Compared to Figure
    \ref{fig_test_success_simulation_no_numin}, even for low sample sizes, test
    success $\le\alpha/2=2.5\%$ for $|\olD| \le \Delta = 1\%$ is ensured. To our
    surprise, the partitioned equivalence test can even be closer to the
    theoretical success probability than the original equivalence test: we use
    $\numin$ for both tests, but only in the non-partitioned test, the threshold
    is never triggered, since safe and unsafe videos are mixed up with $\nu$
    being around $10\%$. Also, the partitioned equivalence test has a larger
    $\nrec$ to draw videos from.}
    \label{fig_test_success_simulation_numin_3}
\end{figure}

\begin{figure}
    \setlength{\unitlength}{1cm}
    \includegraphics[page=1,width=0.95\textwidth]{%
        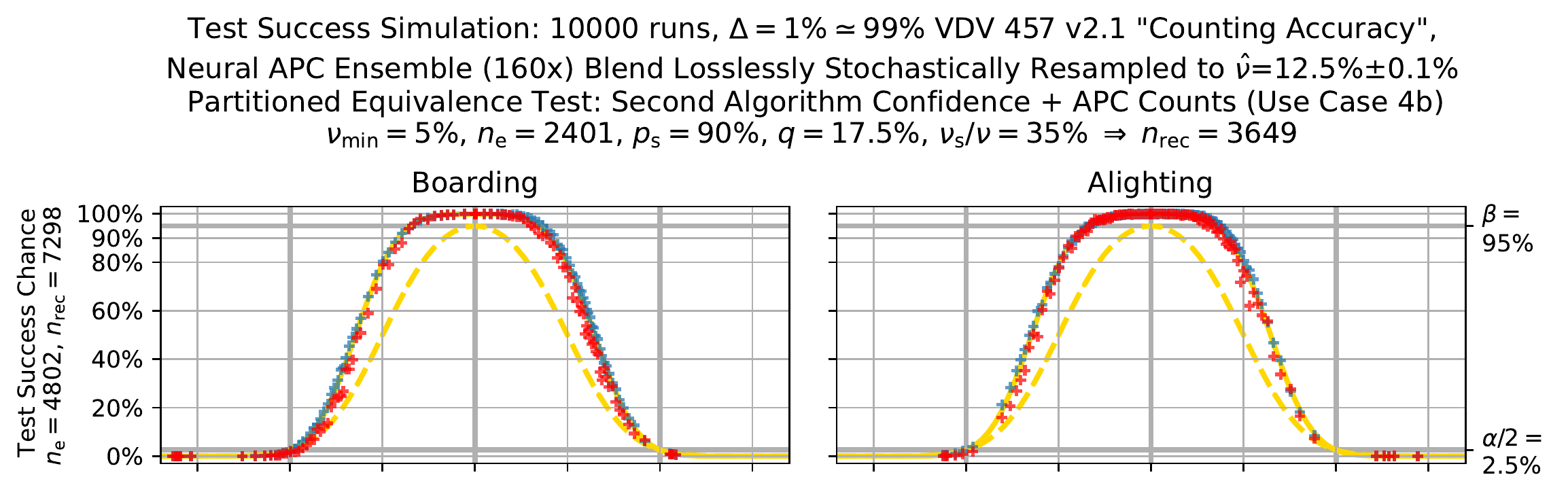}
    \includegraphics[page=1,width=0.95\textwidth]{%
        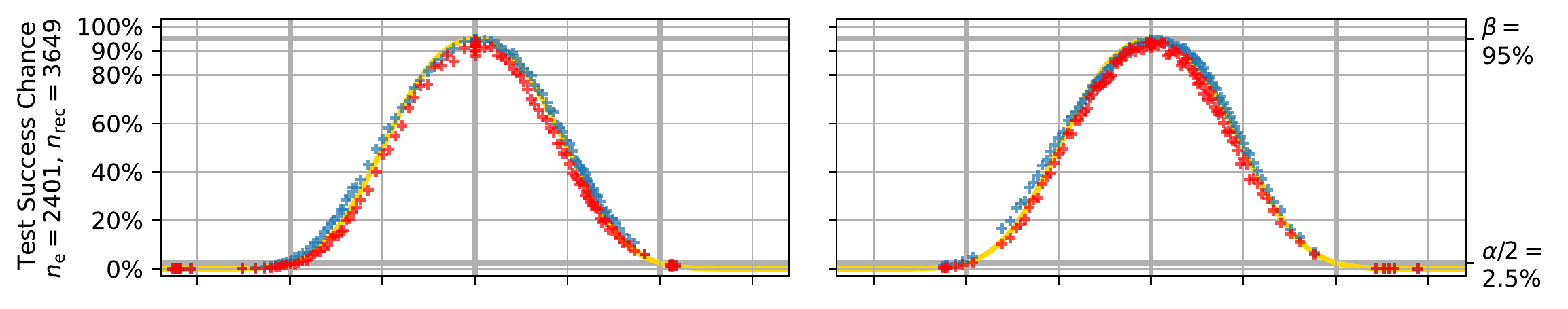}
    \includegraphics[page=1,width=0.95\textwidth]{%
        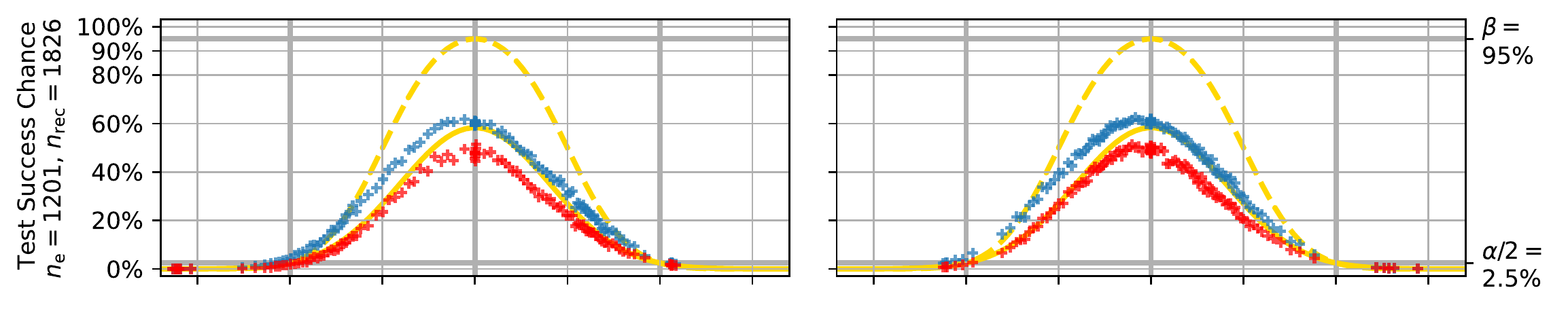}
    \includegraphics[page=1,width=0.95\textwidth]{%
        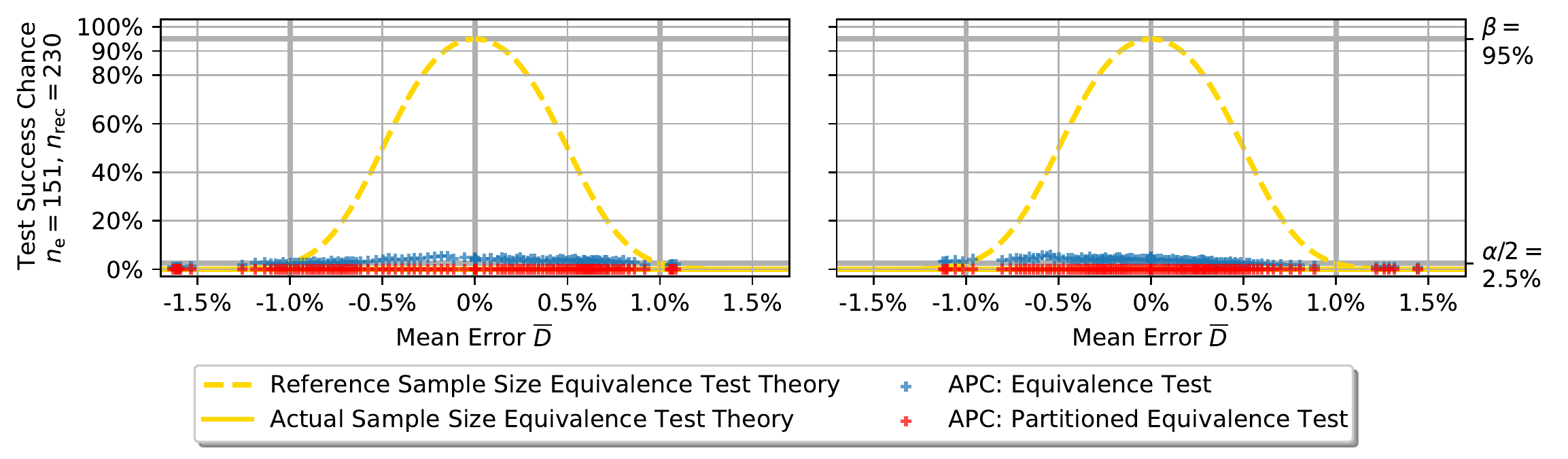}
    \centering
    \caption{Test Success Simulation, comparing the equivalence test with the
    partitioned equivalence test, $\numin=3\%$. As in Figure
    \ref{fig_test_success_simulation_numin_3}, test success $\le\alpha/2=2.5\%$
    for $|\olD| \le \Delta = 1\%$ is ensured. However, $\numin=5\%$ penalizes
    the partitioned equivalence test much stronger than $\numin=3\%$, increasing
    costs or increasing the manufacturers risk. Therefore, we suggest
    $\numin=3\%$.}
    \label{fig_test_success_simulation_numin_5}
\end{figure}

This section consists of various test success simulations, compare Figures
\ref{fig_test_success_theory}, \ref{fig_test_success_simulation_no_numin},
\ref{fig_test_success_simulation_numin_3} and
\ref{fig_test_success_simulation_numin_5}.

\end{appendix}

\end{document}